\documentclass[usenatbib,useAMS]{mnras}

\usepackage{graphics,epsf}
\usepackage{amsmath}                
\usepackage{amsfonts}               
\usepackage{amssymb}                
\usepackage{epsfig}                 
\usepackage{graphicx}
\usepackage{comment}

\usepackage{xcolor}
\definecolor{redak}{rgb}{0.9,0.15,0.05}

\def \kms{~\rm{km~s^{-1}}}

\def \msyr{~\rm{M_{\odot}}~\rm{yr^{-1}}}

\def \K{~\rm{K}}

\def \AU{~\rm{AU}}

\def \days{~\rm{days}}

\def \rmModot{~\rm{M_{\sun}}}
\def \rmRodot{~\rm{R_{\sun}}}
\def \rmLodot{~\rm{L_{\sun}}}

\title[Accretion in Massive Colliding Wind Binaries]{Accretion in massive colliding wind binaries and the effect of wind momentum ratio}

\author[A. Kashi et al.]{
Amit Kashi$^{1,2}$\thanks{E-mail: \href{mailto:kashi@ariel.ac.il}{kashi@ariel.ac.il}},
Amir Michaelis$^{1}$
and
Yarden Kaminetsky$^{1}$
\\
$^{1}$Department of Physics, Ariel University, Ariel, 4070000, Israel\\
$^{2}$Astrophysics, Geophysics and Space Science (AGASS) Center, Ariel University, Ariel, 4070000, Israel\\}

\date{Accepted XXX. Received YYY; in original form ZZZ}

\pubyear{2022}

\begin{document}
\label{firstpage}
\pagerange{\pageref{firstpage}--\pageref{lastpage}}
\maketitle

\begin{abstract}
We carry out a numerical experiment of ejecting winds in a massive colliding wind binary system, and quantifying the accretion onto the secondary star under different primary mass loss rates.
We set a binary system comprising a Luminous Blue Variable (LBV) as the primary and a Wolf-Rayet (WR) star as the secondary, and vary the mass loss rate of the LBV to obtain different values of wind momentum ratio $\eta$.
Our simulations include two sets of cases: one where the stars are stationary, and one that includes the orbital motion.
As $\eta$ decreases the colliding wind structure moves closer to the secondary. We find that for $\eta \lesssim 0.05$ the accretion threshold is reached and clumps which originate by instabilities are accreted onto the secondary.
For each value of $\eta$ we calculate the mass accretion rate and identify different regions in the $\dot{M}
_{\rm acc}$ -- $\eta$ diagram.
For $0.001 \lesssim \eta \lesssim 0.05$ the accretion is sub- Bondi-Hoyle-Lyttleton (BHL)
and the average accretion rate satisfies the power-law
$\dot{M}_{\rm acc} \propto \eta^{-1.73}$ for static stars.
The accretion is not continuous but rather changes from sporadic to a larger duty cycle as $\eta$ decreases.
For $\eta\lesssim0.001$ the accretion becomes continuous in time and the accretion rate is BHL, up to a factor of 0.4--0.8.
The simulations that include the orbital motion give qualitatively similar results, with the steeper power law $\dot{M}_{\rm acc} \propto \eta^{-1.86}$ for the sub-BHL region and lower $\eta$ as an accretion threshold.
\end{abstract}


\section{INTRODUCTION}
\label{sec:intro}

The evolution of very massive stars is quite different from that of low mass stars, and some parts of it are not yet fully understood \citep[e.g.,][]{Smartt2009,Georgyetal2017,Farrelletal2022,EldridgeStanway2022}.
Massive stars involve physical processes that are rarely seen in low mass stars, most notably strong winds and eruptive outbursts \citep[e.g.,][]{Hegeretal2000, KudritzkiPuls2000, Pulsetal2008, Maeder2009, DavidsonHumphreys2012, Vink2015, Kashietal2016}.
One of the quantities that is identified with these stars is their high mass loss rate winds. The wind gets more intense as the stars evolve off the main-sequence (MS) \citep[e.g.,][]{Langer2012, Owockietal2015, Vink2015}.

Perhaps the most interesting phenomena observed in massive stars are related to the fact that most of them reside in binary systems.
\cite{Masonetal2009} found a companion fraction of 75\% for O-stars in Milky Way star clusters.
\cite{Sanaetal2012} estimated that 70\% of O stars have a companion close enough that they will transfer mass during part of their evolution, and that about a third of O stars have a companion so close to the extent that they will merge.
Massive stars interact with their companion star (which is in many cases a massive star by itself) in a way that can affect both stars through mass transfer, common envelope evolution, tidal forces that exert mixing, and irradiation \citep[e.g.,][]{Smith2014,Eldridgeetal2015,DeMarcoIzzard2017,
Eldridge2017,Schroderetal2021,Zapartasetal2021}.
The result of these effects is the alteration of the single star evolution scheme when a close companion is present.
Some evolutionary stages, most notably the LBV and WR-star stages, are now thought to be reached mainly through binary evolution with mass exchange rather than through a single star channel  \citep[e.g.,][]{Eldridgeetal2008,SmithTombleson2015,Mahyetal2022}.
It is therefore important to give attention to binary interactions when studying massive stars.

When a massive binary system has two stars that both eject winds, the two winds collide and create a structure referred to as the colliding wind structure (CWS) \citep{Stevensetal1992,Usov1992,EichlerUsov1993}. These papers draw the following geometric representation of the colliding winds problem:
The CWS is classically defined as an axis-symmetric conical-like shape with increasing thickness further away from its apex. Its hypothetical symmetry line if omitting the orbital motion is a line connecting the two stars, but in fact no symmetry line exists when orbital motion is included as the CWS curves. Still, the apex in most cases is close to the theoretical symmetry line. The CWS is divided by a contact discontinuity with shock waves on both sides. The shocked gas is assumed to flow away asymptotically along the sides of the cone.
Namely, in the classical case the shocked gas resides between each of the two shocks and the contact discontinuity \citep{Usov1992}.
This picture is true close to the stars, which is the region we simulate and discuss here. Further away the curved CWS can collide with itself and the shape looks different, in most cases resembling a spiral. 

The main parameter that determines the shape of the contact discontinuity is the momentum ratio of the two winds
\begin{equation}
\eta=\frac{\dot{M_2} v_2}{\dot{M_1} v_1},
\label{eq:eta}
\end{equation}
where $\dot{M_2}$ and $v_2$ are the mass loss rate and the velocity of the wind of the secondary, and $\dot{M_1}$ and $v_1$ are the same for the primary.
The primary is considered to be the star that has the wind with the larger momentum; i.e. larger $\dot{M} v$. In WR--O binary systems the WR is thus the primary, and in LBV--WR the LBV is the primary.
The structure can take a conical-like shape or a spiral shape depending on the momentum ratio of the winds and the ratio between the winds velocity and the orbital velocity.

Typical values for the mass-loss rate for O-stars are $\dot{M}_{\rm O} \sim 10^{-10}$--$10^{-5} \msyr$ depending on their evolutionary stage 
and their terminal velocity range is  and $v_{\rm \infty,O} \sim 50$--$300 \kms$ \citep[e.g.,][]{Pauldrachetal1986, MullerVink2008, Marcolinoetal2009, Vink2015, Kobulnickyetal2019}.
For WR stars, the ranges are $\dot{M}_{\rm WR} \sim (0.8$--$8) \times 10^{-5} \msyr$ and $v_{\rm \infty,WR} \sim (1$--$5) \times 10^3 \kms$ \citep[e.g.,][]{Usov1992,Crowther2007,Pulsetal2008}. For LBVs the mass loss rates are within $\dot{M}_{\rm LBV} \sim 10^{-6}$--$10^{-3} \msyr$ during quiescence and up to $\sim 1 \msyr$ rate during giant eruptions, and the terminal velocities range is $v_{\rm \infty,LBV} \sim 100$--$500 \kms$ \citep[e.g.,][]{Leithereretal1997, DavidsonHumphreys2012, Vink2015, WeiseBomans2020}.

For O-stars and WR-stars the wind is optically thick and can maintain a steady-state, having the gas coupled to the radiation by bound-bound opacity originated from a known set of spectral lines \citep{Pulsetal2008, Owocki2010, Smith2014}. 
The mass loss fits well with the CAK line-driving model \citep{Castoretal1975}. 
LBVs can lose mass via ordinary line-driving while in quiescent phase, but can have strong continuum driven super-Eddington eruptions \citep{vanMarleetal2008, vanMarleetal2009, Quataertetal2016}.
Clumping can also play a role and affect the mass loss rates of massive stars by a factor of a few \citep{LepineMoffat1999, NugisLamers2000, Pulsetal2006,  Oskinovaetal2007, Sundqvistetal2010, Hainichetal2014}.

New massive colliding wind systems are being observed and known ones monitored, and even those studied for many years reveal interesting new findings.
One recent example is the prototype colliding wind binary WR~140, for which \cite{Zhekov2021} studied RXTE observations and found that a standard colliding wind model with smooth winds does not match the X-ray line profiles. They suggested that adding clumps to the WR wind can solve the discrepancy, and concluded that the clumps are efficiently dissolved in the colliding wind region when the stars are near apastron but not at periastron (the system is highly eccentric $e\approx 0.9$).
\cite{Eatsonetal2022} found that the dust production of WR~140 depends on the orbit with a higher rate close to periastron passage due to formation of clumps by instabilities induced by cooling.

A few massive colliding wind systems were detected and monitored outside of the galaxy, such as Melnick 34 \citep{Pollocketal2018,Tehranietal2019}, Melnick 33Na \citep{Bestenlehneretal2022} and V.~R~144 \citep{Shenaretal2021} in the Large Magellanic Cloud (LMC) 30~Doradus region.
\cite{Williamsetal2021} observed the 2018 dust formation episode in the WR star HD~38030 in the LMC, followed by detection of absorption lines in the next years, indicating an O-star companion in a colliding wind system.
In the Small Magellanic Cloud (SMC) the most notable colliding wind system is HD~5980.
This multiple system contains a 19.3 day eclipsing binary system comprising an LBV primary, which has had LBV eruptions, the most recent of which was in 1993--1994, and possibly an earlier eruption in 1960, and a WR secondary \citep{Koenigsbergeretal2014, Nazeetal2018, Hillieretal2019}.
\cite{Garofalietal2019} found N604-WRX, the first colliding wind binary in M33, composed of a WC4 star and an O-star companion with a long period of $\approx11$ years.

If a massive star undergoes a core collapse SN in a colliding wind system it can have a different lightcurve than a single star SN.
\cite{Pejchaetal2022} calculated the theoretically expected observational signature from such an event. They modeled the SN explosion as it interacts with the CWS and its influence on the shock powered light curve and flash ionization signature, and derived the mass loss rate required to exceed optical luminosities of normal Type~IIp SN.

The colliding winds can be described analytically only when they are adiabatic \citep{Stevensetal1992, Usov1992}. When radiative cooling is introduced, the flow develops instabilities and loses its homogeneity.
Orbital motion, even in a circular orbit, which is not the case for many systems, makes the CWS asymmetric and even wind around itself.
It is therefore natural to use numerical simulations to address the colliding wind problem.
The first 3D simulations of colliding winds were performed for $\gamma^2$~Vel and WR~140 \citep{FoliniWalder2000,FoliniWalder2002,WalderFolini2000,WalderFolini2002,WalderFolini2003}.
These simulations showed the formation of the pinwheel structure of the colliding winds, accounted for the variability obtained in X-rays, and demonstrated the formation of instability and clumping in the winds.

Accretion onto compact objects or point-sources that do not have their own winds has been simulated in many studies first in 2D and later in 3D \citep[e.g.,][]{Sokeretal1986,Ishiietal1993,Ruffertetal1994,Nagaeetal2004}.
Accretion \textit{against wind} is a more rare event and also more difficult to obtain in simulations.
\cite{Akashietal2013} simulated the very massive colliding wind binary $\eta$~Car and found that dense clumps are formed by instabilities in the shocked primary wind as the winds collide. They also found that as these clumps flow towards the secondary they cannot be decelerated by the ram pressure of the secondary wind and hit the secondary.
In \cite{Kashi2017} we studied accretion during periastron passage in $\eta$ Carinae. Using 3D hydrodynamic simulations we showed that the smooth stellar winds collide and develop instabilities, mainly the non-linear thin shell instability (NTSI;\citealt{Vishniac1994}),
and form filaments and clumps. Shortly before periastron passage the dense filaments and clumps flow towards the secondary as a result of its gravitational attraction, and reach the zone where its wind is injected. 
In \cite{Kashi2019} we studied four methods for treating accretion and the response of the accretor to the incoming wind.
In \cite{Kashi2020} we studied accretion in the binary system HD 166734 that contains  
O7.5If primary and O9I(f) secondary. The orbital period is $P \simeq 34.5$ days, the eccentricity is $e \simeq 0.62$ and the wind momentum ratio is $\eta \simeq 0.32$. The winds were highly unstable as they were radiative from both sides, creating long non-linear thin shell instability fingers. We obtained accretion for a long duration of the orbital period, that sums up to $\dot{M}_{\rm acc} \simeq 1.3 \times 10^{-8} \rmModot$ each cycle.

Though there are advanced numerical simulations studying colliding winds, most of them are focused on one specific binary system with its particular conditions (properties of the stars and winds, orbital parameters). There is no set of \emph{general} simulations that ran over a range of each of the parameters and isolated the influence of each physical effect.
While obtaining such a set requires a very large number of simulations, focusing on specific links between parameters through a limited set of simulations is a more reachable goal.
In this paper we show that this is possible through a sweep of specific parameters in a set of simulations. The parameter swept here is the mass loss rate of the primary, which is a proxy to the momentum ratio $\eta$. We change it in each simulation while the rest of the parameters do not vary. We focus on the influence of $\eta$ on the accretion rate and properties onto the companion.

Some of the early results reported in this paper were presented in \cite{KashiMichaelis2021}. 
In section \ref{sec:simulation} we describe the numerical simulation.
Our results  are presented in section \ref{sec:results} and discussed in section \ref{sec:discussion}.
Our summary is given in section \ref{sec:summary}.

\begin{table*}
\caption{A summary of simulations and results for our static simulations (with no orbital motion). For all simulations $a=6 \AU$, $M_1 =80 \rmModot$, $M_2 =20 \rmModot$, $v_{1,\infty} = 500 \kms$, $v_{2,\infty} = 3000 \kms$ . Numbers in parenthesis indicate power of 10. Run 1 is the fiducial run with primary mass loss rate $\dot{M}_{1,f}=3 \times 10^{-4} \msyr$ (accretion does not occur for the fiducial run). Runs 2--10 belong to region (ii) and runs 11--14 to region (iii).}
\begin{tabular}{lcccccccccc}

\hline\hline

Parameter                    				        	& Run 1      & Run 2      & Run 3       & Run 4       & Run 5      & Run 6      & Run 7      & Run 8       & Run 9       & Run 10     \\
\hline         
$\dot{M}_1/\dot{M}_{1,f}$       				        & 1          & 4          & 5           & 8           & 10          & 15         & 20         & 40		   & 60          & 80	      \\  
$\dot{M}_1 (\rm{M_{\odot}}~\rm{yr^{-1}})$               & 3(-4)      & 1.2(-3)    & 1.5(-3)     & 2.4(-3)     & 3(-3)       & 4.5(-3)    & 6(-3)      & 0.012      & 0.018       & 0.024      \\
$\eta$                       				        	& 0.2        & 0.05       & 0.04        & 0.025       & 0.02        & 0.0133     & 0.01       & 5(-3)	   & 3.33(-3)    & 2.5(-3)	  \\
$\dot{M}_{\rm acc, av} (\rm{M_{\odot}}~\rm{yr^{-1}})$   & 0          & 6.1(-9)    & 1.4(-8)     & 1.9(-8)     & 4.9(-8)     & 6.7(-8)     & 1.5(-7)   & 4.2(-7)    & 9.4(-7)     & 1.1(-7)    \\ 
$D_{\rm acc}$                                           & 0          & 0.041      & 0.064       & 0.095       & 0.131       & 0.205      & 0.413      & 0.819      & 0.956       & 0.976      \\

\hline\hline
Parameter                    			        		& Run 11     & Run 12     & Run 13      & Run 14    &   &   &   &   &   & \\
\hline         
$\dot{M}_1/\dot{M}_{1,f}$       				        & 160        & 320        & 640         & 1280      &   &   &   &   &   & \\
$\dot{M}_1 (\rm{M_{\odot}}~\rm{yr^{-1}})$               & 0.048      & 0.096      & 0.192       & 0.384     &   &   &   &   &   & \\
$\eta$                       			        		& 1.25(-3)   & 6.25(-4)   & 3.125(-4)   & 1.5625(-4)&   &   &   &   &   & \\
$\dot{M}_{\rm acc, av} (\rm{M_{\odot}}~\rm{yr^{-1}})$   & 4.6(-6)    & 7.4(-6)    & 1.2(-5)     & 2.2(-5)   &   &   &   &   &   & \\
$D_{\rm acc}$                                           & 1          & 1          & 1           & 1         &   &   &   &   &   & \\

\end{tabular}
\label{table:parameters} 
\end{table*}

\begin{table*}
\caption{A summary of simulations and results for runs with orbital motion. The parameters are similar to Table \ref{table:parameters} but the period is $P =536.68 (\rm{days})$ and the eccentricity is $e=0$. Numbers in parenthesis indicate power of 10.}
\begin{tabular}{lcccccccccc}

\hline\hline

Parameter                    				        	& Run 1R     & Run 2R      & Run 3R     & Run 4R      & Run 5R      & Run 6R     & Run 7R      & Run 8R    & Run 9R      & Run 10R   \\
\hline         
$\dot{M}_1/\dot{M}_{1,f}$       			        	& 1          & 4          & 5           & 8           & 10          & 15         & 20         & 40		   & 60          & 80	      \\  
$\dot{M}_1 (\rm{M_{\odot}}~\rm{yr^{-1}})$               & 3(-4)      & 1.2(-3)    & 1.5(-3)     & 2.4(-3)     & 3(-3)       & 4.5(-3)    & 6(-3)      & 0.012      & 0.018       & 0.024      \\
$\eta$                       					        & 0.2        & 0.05       & 0.04        & 0.025       & 0.02        & 0.0133     & 0.01       & 5(-3)	   & 5(-3)       & 2.5(-3)	  \\
$\dot{M}_{\rm acc, av} (\rm{M_{\odot}}~\rm{yr^{-1}})$   & 0          & 0          & 8.6(-9)     & 1.0(-8)     & 3.1(-8)     & 7.2(-8)    & 1.2(-7)    & 4.2(-7)    & 6.8(-7)     & 1.1(-6)    \\ 
$D_{\rm acc}$                                           & 0          & 0          & 0.051       & 0.075       & 0.124       & 0.153      & 0.254      & 0.520      & 0.798       & 0.855      \\

\hline\hline
Parameter                              					& Run 11R    & Run 12R     & Run 13R    & Run 14R    &   &   &   &   &   & \\
\hline         
$\dot{M}_1/\dot{M}_{1,f}$                				& 160        & 320        & 640         & 1280       &   &   &   &   &   & \\
$\dot{M}_1 (\rm{M_{\odot}}~\rm{yr^{-1}})$               & 0.048      & 0.096      & 0.192       & 0.384      &   &   &   &   &   & \\
$\eta$                       			         		& 1.25(-3)   & 6.25(-4)   & 3.125(-4)   & 1.5625(-4) &   &   &   &   &   & \\
$\dot{M}_{\rm acc, av} (\rm{M_{\odot}}~\rm{yr^{-1}})$   & 4.5(-6)    & 7.6(-6)    & 1.2(-5)     & 2.2(-5)    &   &   &   &   &   & \\
$D_{\rm acc}$                                           & 1          & 1          & 1           & 1          &   &   &   &   &   & \\

\end{tabular}
\label{table:parametersR} 
\end{table*}

\section{THE NUMERICAL SIMULATIONS}
 \label{sec:simulation}

We aim to test the effect of enhanced primary wind on the amount of accreted gas onto the secondary.
We neutralize the effect of orbital motion, in order to isolate the effect of the primary mass loss rate.
The effect of enhancing the primary mass loss rate is equivalent to decreasing the momentum ratio.

We run the numerical experiment using the hydrodynamic code \textit{FLASH}, described in \cite{Fryxell2000}.
Our 3D Cartesian grid extends over $x=(-8,8) \AU$, $y=(-8,8) \AU$ and $z=(-4,4) \AU$. The grid is centered around the secondary star, and the orbital plane is on the $x$--$y$ plane.
The grid has 5 levels of refinement that are centered on the secondary. The largest cell has a side of $\simeq 18 \rmRodot$ and the smallest has a side of $\simeq 1.1 \rmRodot$. 
The apex of the colliding winds has the second finest resolution, with cell sides of $\simeq 2.2 \rmRodot$.
With this configuration we have high resolution on the colliding winds close to their apex.
To solve the hydrodynamic equations we use the \textit{FLASH} version of the split piece-wise parabolic method (PPM) solver \citep{ColellaWoodward1984}, an extension of the Godunov method, which solves a Riemann
problem and allows resolving shocks.
A multigroup diffusion approximation is used for solving the radiative transfer equation and updating the energy.
We also include radiative cooling from \cite{SutherlandDopita1993}, which is necessary in modeling colliding winds governed by instabilities. The temperature is limited from below to $10\,000\K$.

We set up a stationary binary system with binary separation $a=6 \AU$ which is within the range of known similar systems and fits within the grid to allow us to capture the CWS as well as the downstream behind the secondary. The stellar masses ${M_1 = 80 \rmModot}$ and $M_2=20 \rmModot$, also typical values. The effective temperature of the primary is $T_{\rm eff,1} = 20\,000 \K$, its radius is taken to be $R_1=100\rmRodot$ and the luminosity is $L_1 \simeq 1.4\times 10^6 \rmLodot$. The secondary is hotter and smaller with $T_{\rm eff,2} = 40\,000 \K$, $R_2=20\rmRodot$, and $L_2 \simeq 9\times 10^5 \rmLodot$.
The composition used for both stars is appropriate for an evolved star with CNO-processed material and the He abundance is 50 percent. This composition is also used for calculating the opacity, for which an \textit{OPAL} model is used \citep{IglesiasRogers1996}.
The primary mass loss rate is $\dot{M}_1 = 3 \times 10^{-4} \msyr$ for the fiducial case (Run 1) and its wind velocity has a terminal value of $v_{1,\infty}=500 \kms$ with radiative acceleration corresponding to $\beta=1$ \citep{Pauldrachetal1986}.
These values are appropriate for an LBV star and its wind \citep[e.g.,][]{Vink2015}.
The interiors of the stars are not part of the simulation. The gravitational field is modeled by two point sources at the centers of both stars.
The secondary mass loss rate is $\dot{M}_2 =10^{-5} \msyr$, and its wind velocity has a terminal value $v_{2,\infty} = 3000 \kms$ with an acceleration parameter $\beta=0.8$.
These values are appropriate for a WR star and its wind \citep[e.g.,][]{Pauldrachetal1986,MullerVink2008,Vink2015}.
The wind is ejected from a narrow shell around each star. Both stars accelerate their winds according to a $\beta$-law. In addition, radiation-transfer also takes place.

\begin{figure}
\centering
\includegraphics[trim= 0.0cm 0.0cm 0.0cm 0.0cm,clip=true,width=0.99\columnwidth]{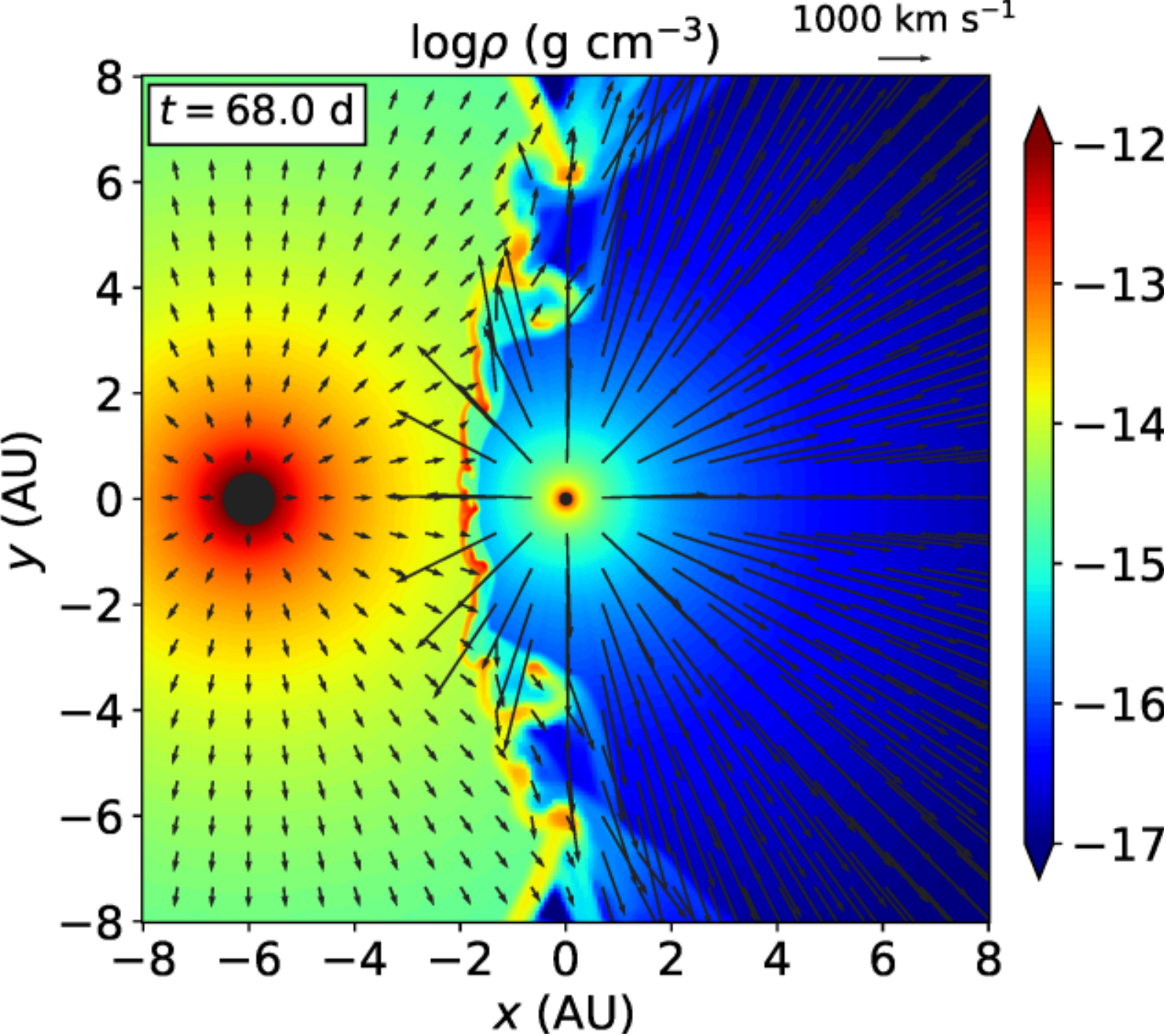}  
\caption{
Density map with velocity vectors showing the density sliced in the orbital plane ($z=0$), for Run 1 ($\eta=0.2$).
The time epoch shown here represents the flow just before the continuation of the experiment, in which we enhance the primary mass loss rate.
The secondary is at the center, marked with a small black circle while the primary, marked with a large black circle, is on the left side.
The two winds are accelerated and collide to form the CWS. The parameters are given in Table \ref{table:parameters}.
The obtained instabilities are the result of radiative cooling. Smaller filaments are formed on the side facing the secondary. For Run 1 there is no accretion.
}
\label{fig:density_smooth_16}
\end{figure}

The momentum ratio for these parameters is
\begin{equation}
\eta= \frac{\dot{M}_2 v_{2,\infty}}{\dot{M}_1 v_{1,\infty}}= 0.2,
\label{eq:eta_fiducial}
\end{equation}
but as the winds collide before reaching their terminal velocity, the effective momentum ratio is different and also varies with location. Nevertheless, to remain consistent with the common definition in the literature we cite the wind momentum ratio as defined by equation (\ref{eq:eta_fiducial}).
Figure \ref{fig:density_smooth_16} shows the test case (that did not lead to accretion onto the secondary).
Since our simulation includes radiative cooling, the colliding wind region develops instabilities along the shocked surfaces.
The place where the winds interact depends on $\eta$. For our fiducial run at the apex the primary and secondary wind reaches $\simeq99\%$  and $\simeq95\%$ of their terminal velocity, respectively.
At other points on the CWS these values are even larger.
For the other simulations, as $\eta$ decreases the intersection moves towards the secondary and therefore the primary is at its terminal velocity while the secondary is not.
%

At $t=68.5 \days$ we increase the mass loss rate of the primary to a larger value, as indicated in Table \ref{table:parameters}.
The enhanced primary wind facing the secondary reaches the apex at $t\simeq82.5$ days.
When it arrives at the colliding wind structure it disrupts its shape and induces stronger instabilities.
The instabilities create dense fingers that penetrate the colliding wind structure and face the secondary. Clumps and filaments are formed due to the instabilities, primarily the NTSI. The secondary gravity then pulls the clumps, confronting the secondary wind that pushes them away. Depending on the dominant force, the clump might get accreted.
We quantify the accretion rate according to the method described in \citep{Kashi2019}.

\section{RESULTS}
\label{sec:results}

We post-process every simulation to derive the quantities that we discussed above.
We obtain a post-processing output every $1/2$ day,
even though our data is calculated in time steps of a few minutes.
We started increasing the primary mass loss rate and checking the amount of accreted mass on the companion as a result.
We tried doubling ($\eta=0.1$) and tripling ($\eta=0.067$) the primary mass loss rate, but this did not lead to any accreted mass, as the secondary wind and radiation was able to push and deflect the incoming primary wind up to this primary mass loss rate.

\begin{figure}
\centering
\includegraphics[trim= 0.0cm 0.0cm 0.0cm 3.0cm,clip=true,width=0.48\textwidth]{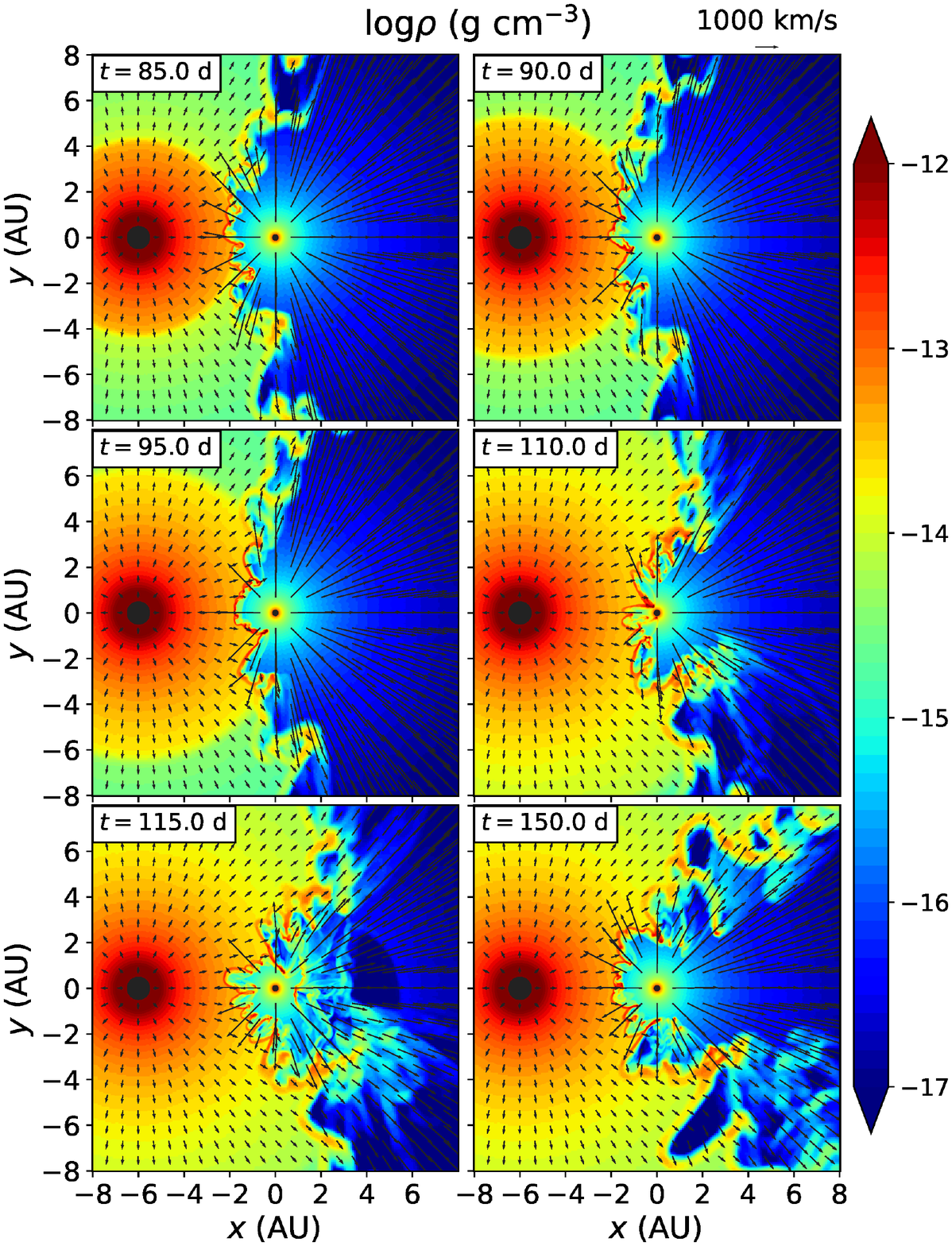}
\caption{
Density maps with velocity vectors showing slices in the orbital plane ($z=0$), for Run 2 ($\eta=0.05$).
Since the numerical experiment is performed on stationary stars with no orbital motion, the slice of the orbital plane is similar to the slice of any other plane around the axis joining the two stars.
The secondary is at the center, marked with a small black circle while the primary, marked with a large black circle, is on the left side.
The two winds are accelerated and collide. The primary then ejects enhanced wind that interacts with the secondary wind at the colliding wind region.
For $\eta>0.05$ we did not obtain accretion, however for this run clumps from the CWS penetrate into the secondary wind and get accreted onto the secondary. Accretion only happened in brief episodes that last $\approx 2 \days$ each, during which a clump is being accreted. In between these episodes there are long periods with no accretion.
}
\label{fig:37}
\end{figure}

When we increased the primary mass loss rate to 4 times the mass loss rate of Run 1, we obtained accretion. This is labeled as Run 2, with $\dot{M}_1=1.2\times 10^{-3} \msyr$, and $\eta=0.05$. Figure \ref{fig:37} shows density maps together with velocity vectors at different times for Run 2.
In each panel the secondary is at the center of the grid, and the primary is to the left.
The primary wind collides with the pre-existing CWS and changes its shape to a smaller opening angle.
The side of the CWS facing the secondary shows strong instabilities and forms dense clumps and filaments. The gravity of the secondary pulls these filaments, and some of them are accreted onto the secondary. The secondary wind tries to flow against the incoming gas and forms bubbles.

Accretion occurred for a single brief episode starting at $t=109.5$ days and lasting $\simeq 2$ days during the transition between the two mass loss rates.
This is shown in the middle-right panel of Figure \ref{fig:37}.
As can be seen, the accretion occurred when the enhanced primary wind pushed the pre-existing CWS towards the secondary, and a clump reached the secondary.
Later accretion events also took place, with durations of $\simeq 2$--$3$ days.
These accretion episodes were also accretion of individual clumps formed at the CWS, that managed to arrive and accrete to the secondary despite its outflow and radiation pressure, that results in radiative breaking of the accreted gas.

We therefore conclude that Run 2 is very close to the conditions where no accretion shall take place for $\eta \lesssim 0.05$, about 4 times the fiducial value of $\dot{M}_{1,f}$ (Run 1).
We find that the average mass accretion rate for Run 2 ($\eta=0.05$) is $\dot{M}_{\rm acc, av} \simeq 6.1 \times 10^{-9} \msyr$. The accretion is intermittent, with irregular intervals and an average duty cycle of $D_{\rm acc} \simeq 0.041$ that we discuss blow (Section \ref{sec:discussion}).

\begin{figure}
\centering
\includegraphics[trim= 0.0cm 0.0cm 0.0cm 3.0cm,clip=true,width=0.48\textwidth]{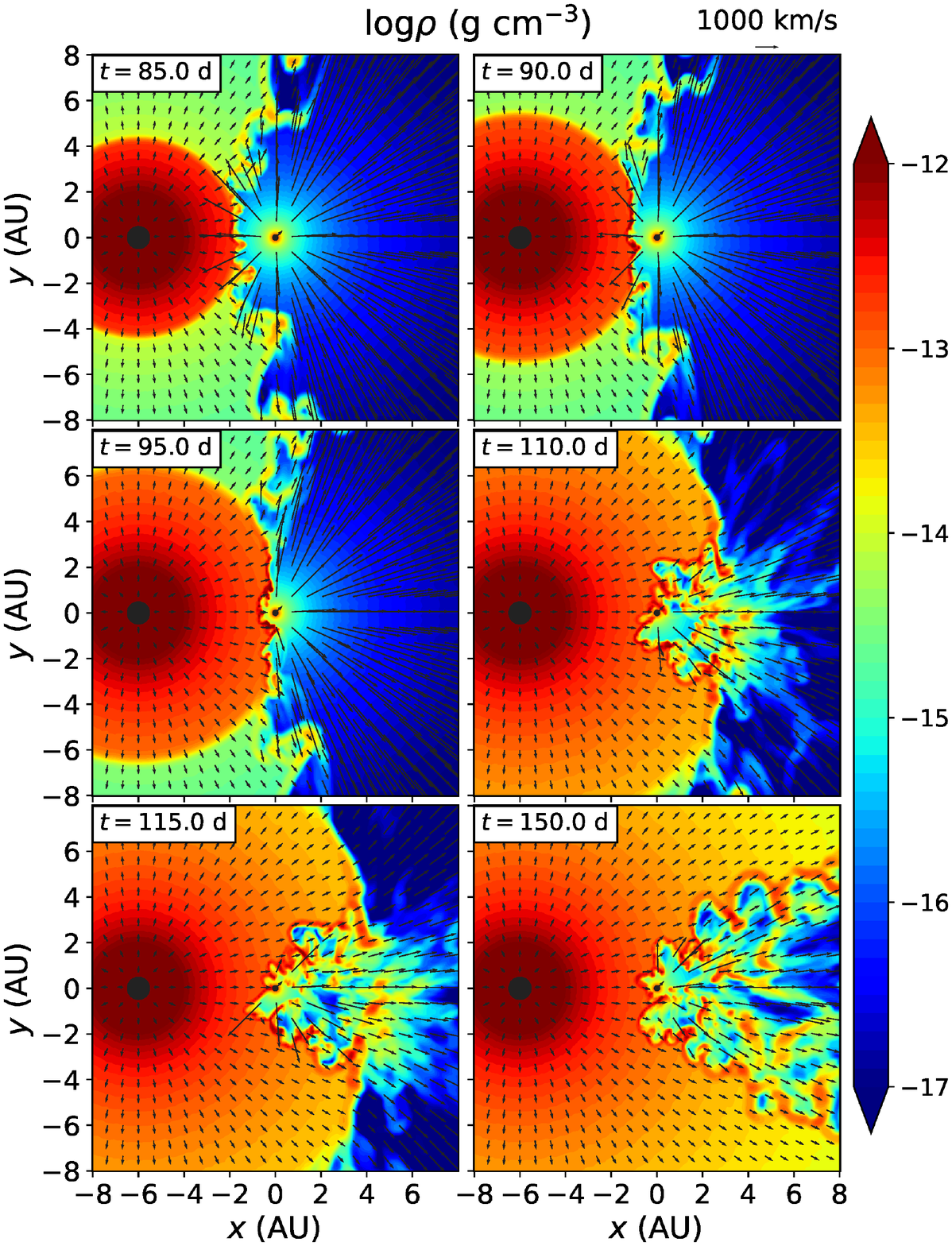}
\caption{
Same as Figure \ref{fig:37}, but for Run 7, with $\eta=0.01$ (the primary mass loss rate is five times larger than in Run 2, and twenty times larger than in Run 1).
The two winds are accelerated and collide. The primary then ejects enhanced wind that interacts with the secondary wind at the colliding wind region, and then penetrates into the secondary wind and accretes onto the secondary.
}
\label{fig:21}
\end{figure}

\begin{figure*}
\centering
\includegraphics[trim= 0.0cm 0.0cm 0.0cm 0.0cm,clip=true,width=0.81\textwidth]{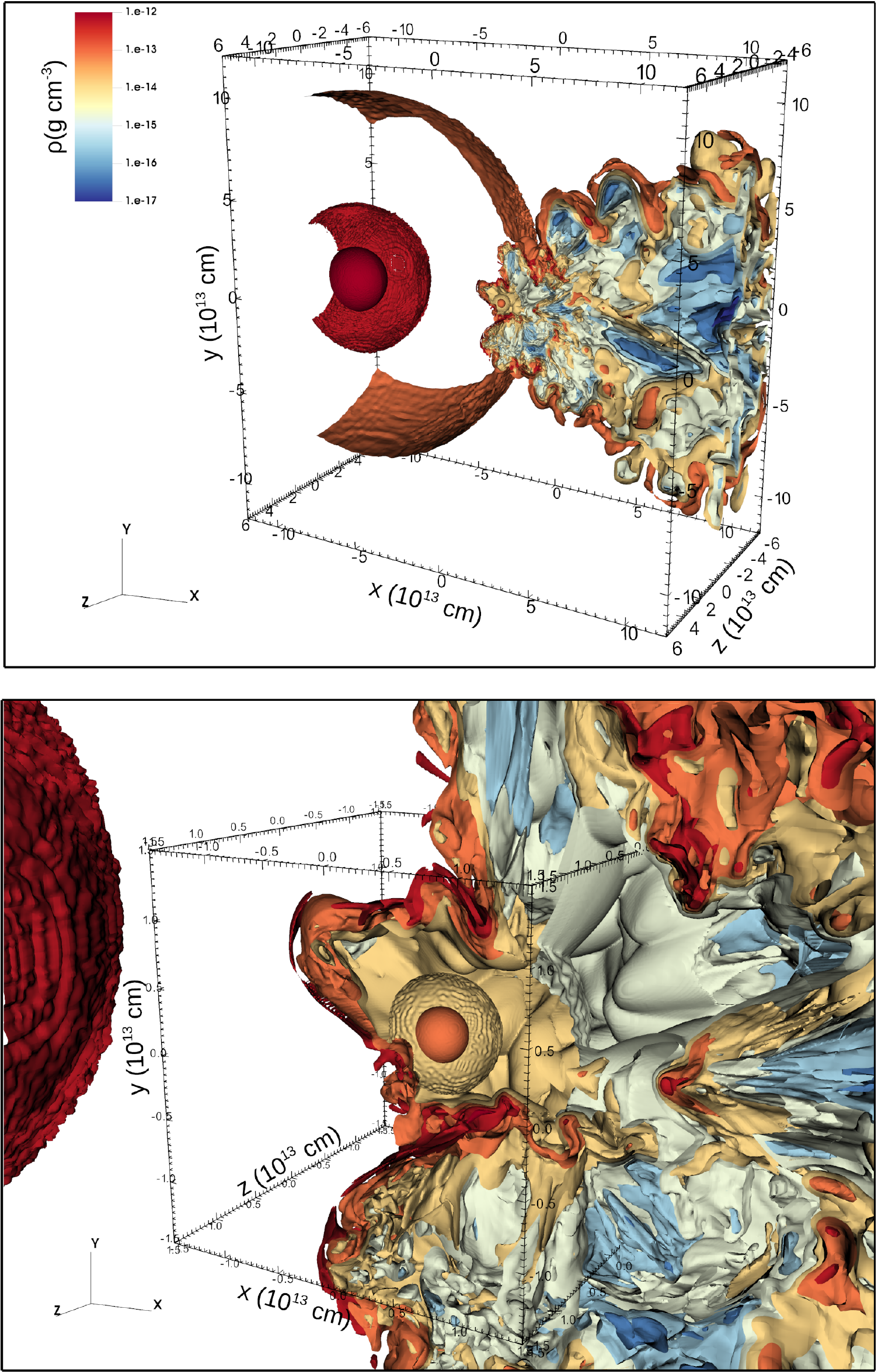}
\caption{
\textit{Upper panel:} A 3D view of iso-density surfaces of Run 7 with $\eta=0.01$ at $t=150$ days (same as lower-right panel in Figure \ref{fig:21}).
The grid is clipped at the orbital plane such that only $z<0$ is shown.
The primary is on the left side and the secondary is at the center. \textit{Lower panel:} A zoomed view near the secondary. The box side is $1 \AU$.
This figure shows finer details of the simulation and specifically the instabilities in the winds.
}
\label{fig:21_300_3d}
\end{figure*}

We then run more runs with lower values of $\eta$. The changes between the runs are gradual. We will hereby discuss some of the runs that feature key points. 
Figure~\ref{fig:21} shows Run 7, for which $\eta=0.01$.
A 3D view of Run 7 at $t=150$ days is shown in Figure \ref{fig:21_300_3d}. The conical-like shape of the CWS is interfered with the large instabilities. During the transition time the inner part of the CWS does not have a clear cavity from the secondary wind but rather has a mixture of material from both winds. Later on, the wind of the secondary manages to reopen the cavity (lower right panel of Figure~\ref{fig:21}). The secondary wind pushes the primary wind and creates bubbles. Between these bubbles dense filaments of gas flow and some of them get accreted onto the secondary while others continue to flow away along the CWS. 
%
%

\begin{figure}
\centering
\includegraphics[trim= 0.0cm 0.0cm 0.0cm 3.0cm,clip=true,width=0.48\textwidth]{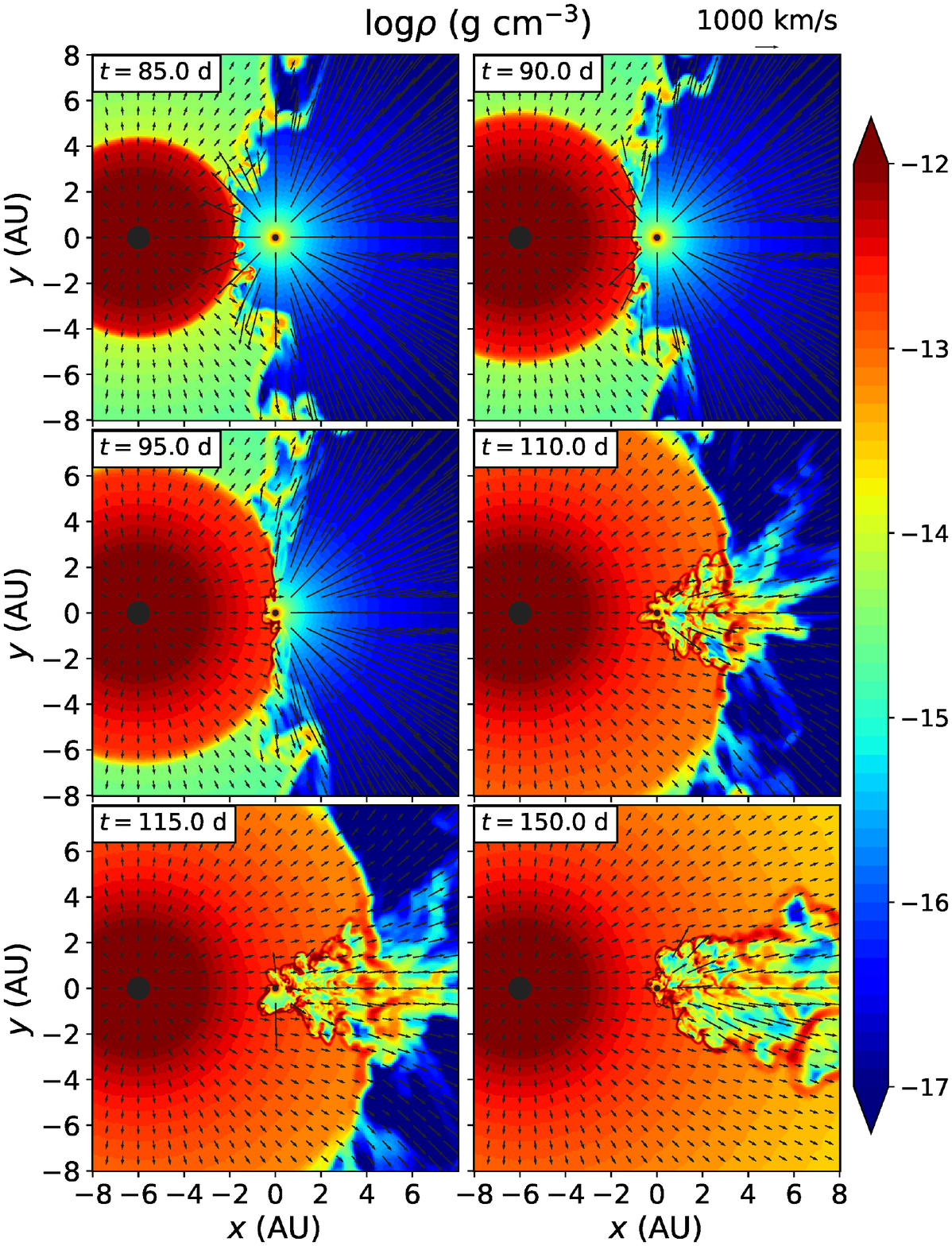}
\caption{
Same as Figure \ref{fig:37}, but for Run 8, with $\eta=0.005$. The CWS is narrower and there is no longer a void behind the secondary with the undisturbed secondary wind.
}
\label{fig:22}
\end{figure}

In Figure~\ref{fig:22} we show Run 8, with $\eta=0.005$. The results are qualitatively similar to Run 7 but a narrower CWS is formed. For this value of $\eta$ there is no longer a distinguishable cavity of secondary wind at any time (not even the small void that appears at late times as seen in Run 7), but instead a cavity filled with turbulent flow composed of the mixture of the two colliding winds.
%

\begin{figure}
\centering
\includegraphics[trim= 0.0cm 0.0cm 0.0cm 3.0cm,clip=true,width=0.48\textwidth]{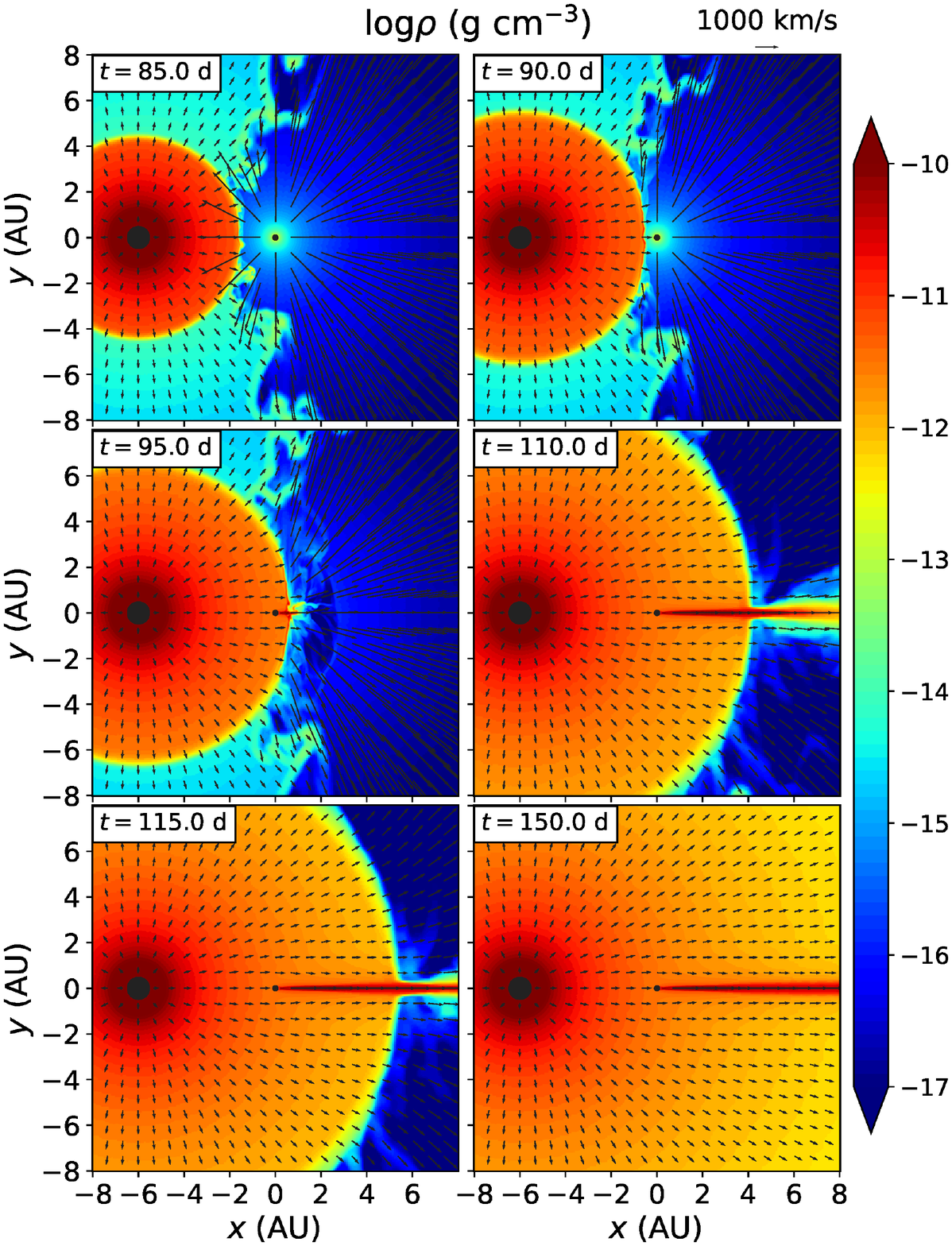}
\caption{
Same as Figure \ref{fig:37}, but for Run 13, with $\eta=3.125\times10^{-4}$. Note the different density scale. The secondary accretes directly from the primary wind from all directions except a narrow solid angle at the side facing away from the primary.
The secondary focuses the primary wind to create a narrow dense column behind the secondary.
}
\label{fig:27}
\end{figure}

Figure \ref{fig:27} presents Run 13, with mass loss rate of $\dot{M}_1 = 0.192 \msyr$, and $\eta=3.125\times10^{-4}$. This high mass loss rate corresponds to a giant LBV eruption, and is an extreme case of a colliding wind binary.
The secondary wind cannot blow against the strong primary wind, and is almost completely suppressed. The secondary accretes directly from the primary wind from all directions except a narrow solid angle at the side facing away from the primary.
The secondary focuses the primary wind to create a narrow dense column behind the secondary, that has therefore higher density than other directions. Some of the accreted mass arrives from behind the star through this column and some directly hits the star from the side facing the primary.
%
%

\begin{figure*}
\centering
\includegraphics[trim= 0.0cm 0.0cm 0.0cm 0.0cm,clip=true,width=0.95\textwidth]{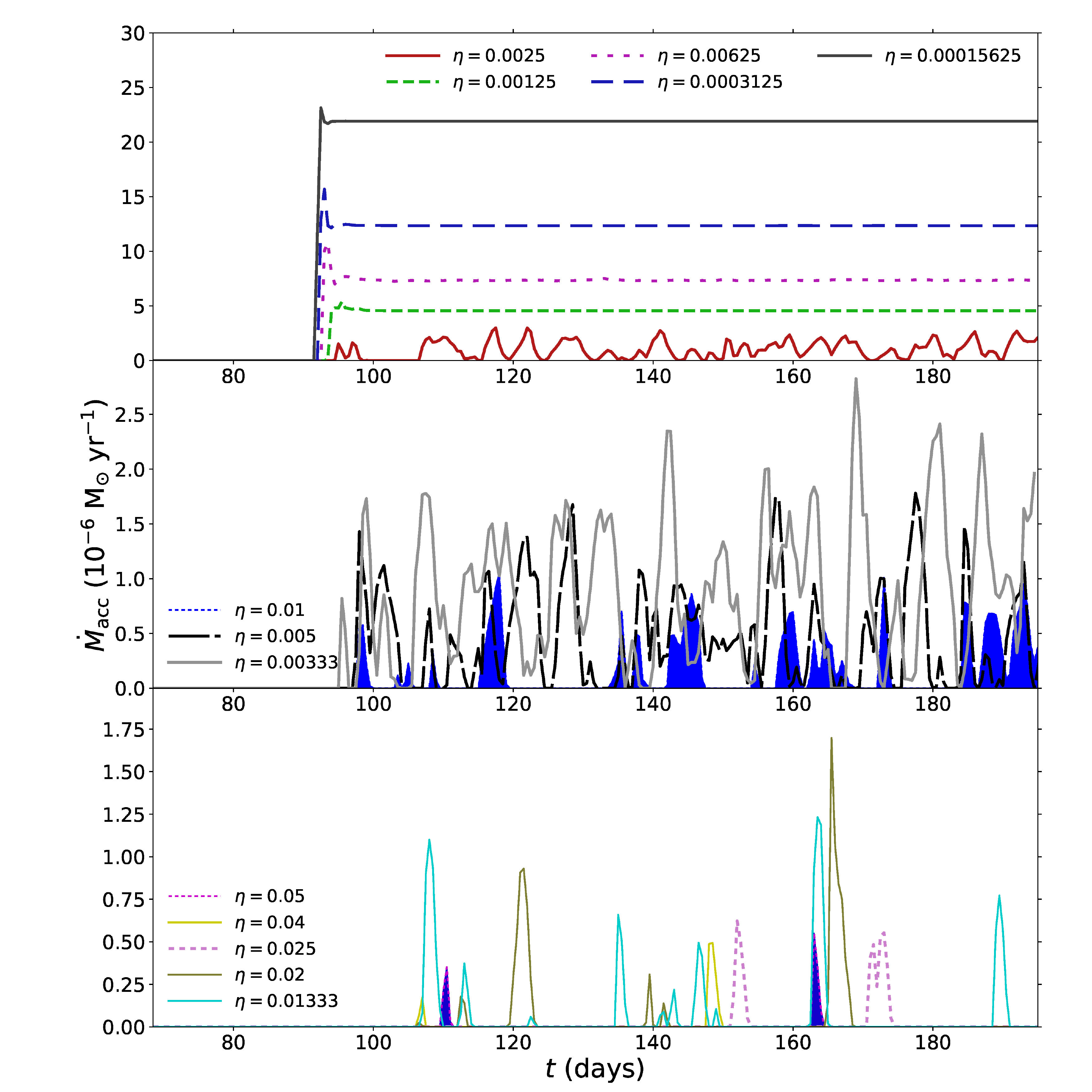} 
\caption{
A panoramic view of the mass accretion rate onto the secondary star for Runs 2--14 (static runs with no orbital motion).
The lines are labeled according to the value of the momentum ratio $\eta$ (Equation \ref{eq:eta}), and the parameters are given in Table \ref{table:parameters}.
The Runs were distributed to 3 panels for clarity. Note the different scale for each panel. Some of the runs were simulated for a longer period of time than shown in the figure for better statistics.
}
\label{fig:3panel_mdot}
\end{figure*}

Figure \ref{fig:3panel_mdot} collects the accretion rate as a function of time for Runs 2--14 (spread over 3 panels to avoid clutter).
The figure shows that the smaller the value of $\eta$ the larger the accretion rate, and the longer the duty cycle of accretion.
We further find that for $\eta\leq0.01$ the smaller the value of $\eta$ the earlier the accretion starts.
%

\begin{figure}
\centering
\includegraphics[trim= 0.0cm 0.0cm 0.0cm 3.0cm,clip=true,width=0.48\textwidth]{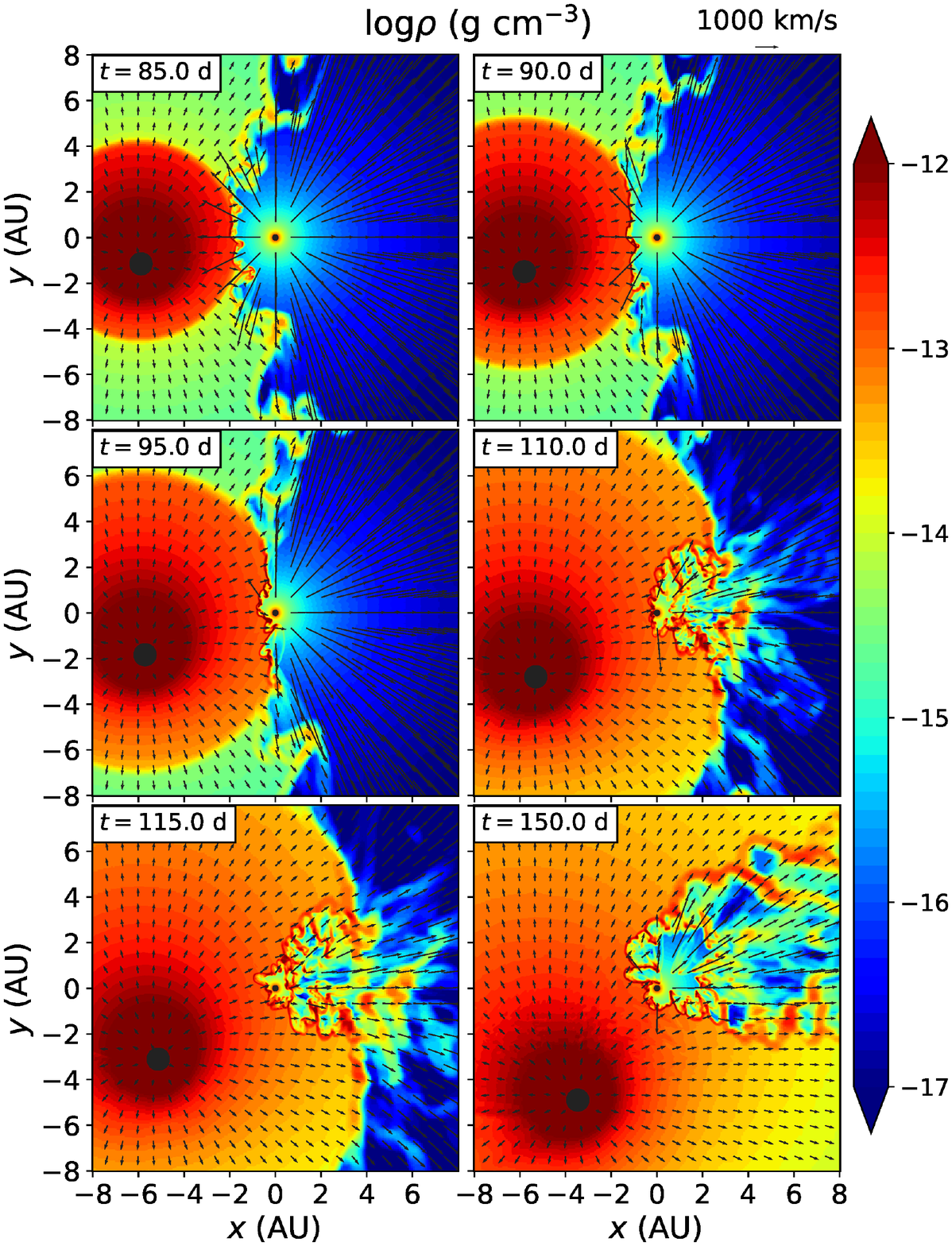}
\caption{
Same as Figure \ref{fig:21}, but for Run 7R, with $\eta=0.01$ and orbital motion. The CWS is rotated and curved. The cavity of the secondary wind contains mixed material from the primary and the secondary. The accretion rate is very close to the non-rotating case.
}
\label{fig:46}
\end{figure}

\begin{figure}
\centering
\includegraphics[trim= 0.0cm 0.0cm 0.0cm 3.0cm,clip=true,width=0.48\textwidth]{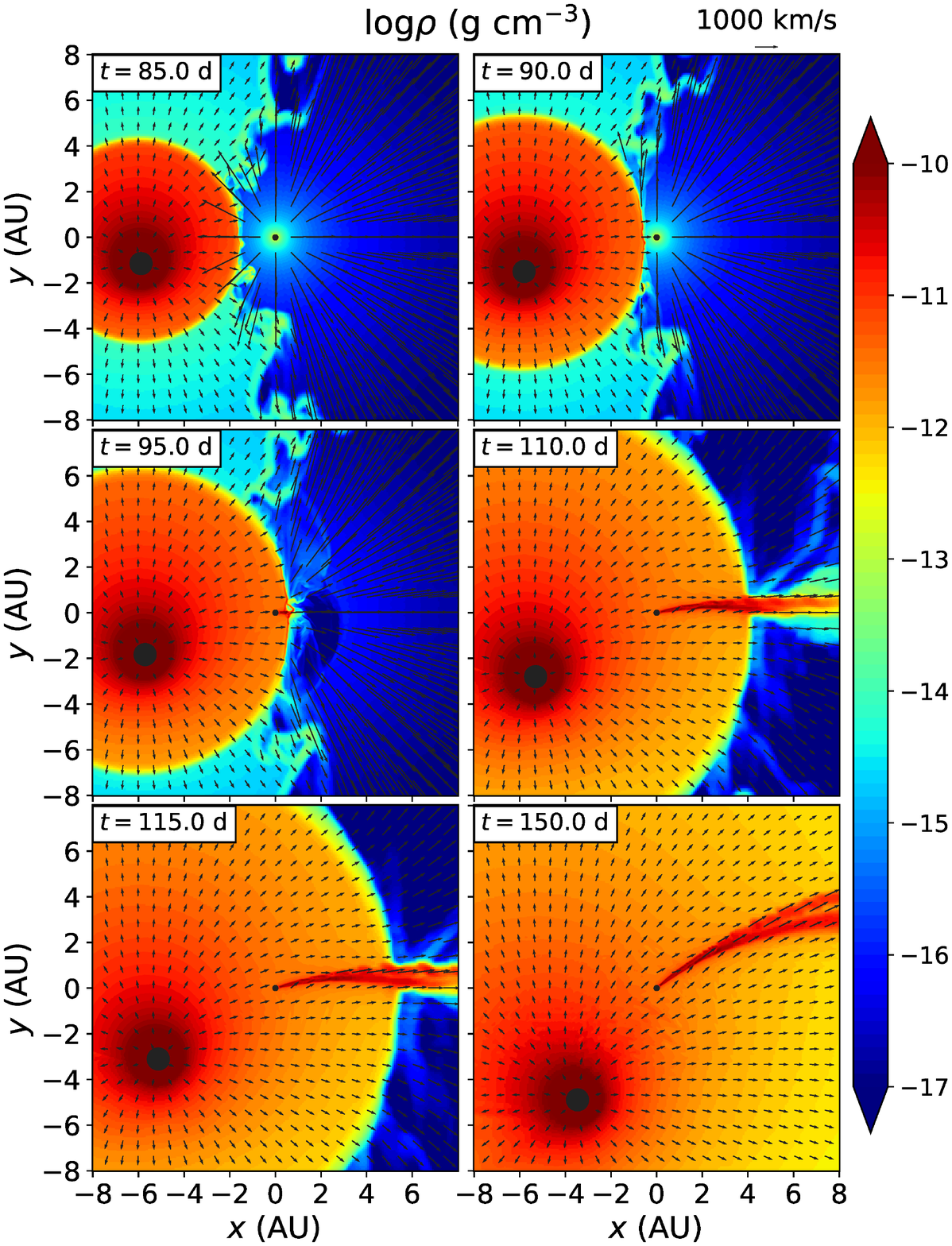}
\caption{
Same as Figure \ref{fig:27}, but for Run 13R, with $\eta=3.125\times10^{-4}$ and orbital motion. The orbital motion creates a curved accretion line.
Accretion is continuous as in the static case. Note the different density scale.
}
\label{fig:53}
\end{figure}

\begin{figure*}
\centering
\includegraphics[trim= 0.0cm 0.0cm 0.0cm 0.0cm,clip=true,width=0.95\textwidth]{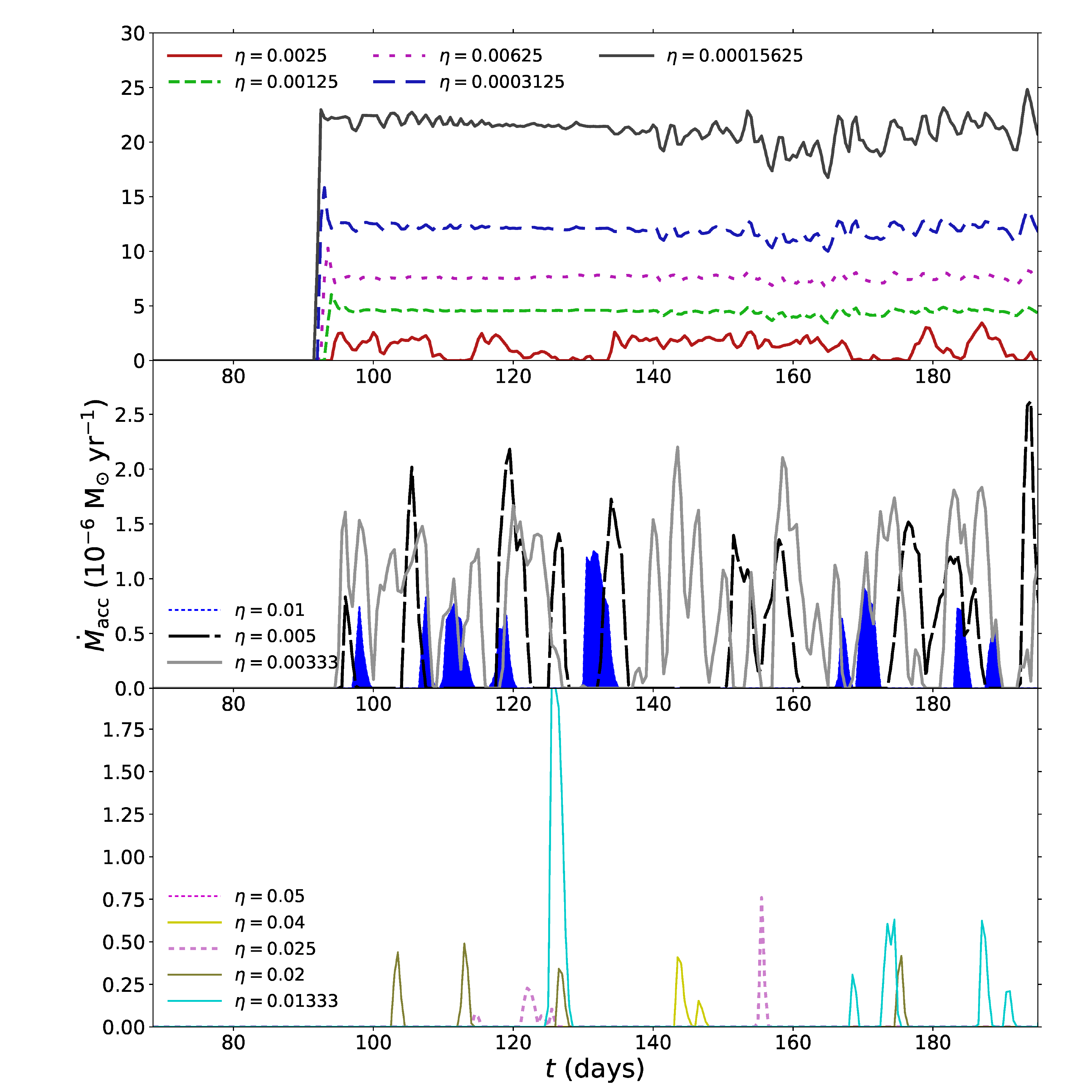} 
\caption{
A panoramic view of the mass accretion rate onto the secondary star for Runs 2R--14R (with orbital motion).
The lines are labeled according to the value of the momentum ratio $\eta$ (Equation \ref{eq:eta}), and the parameters are given in Table \ref{table:parametersR}.
The Runs were distributed to 3 panels for clarity. Note the different scale for each panel.
Note that for $\eta=0.05$ (Run 2R) the line is not shown as there is no accretion.
}
\label{fig:3panel_mdot_orb}
\end{figure*}

We check the effect of adding orbital motion. We take a circular orbit with orbital velocity that satisfies ${v_{\rm orb} \simeq 0.25 v_{1,\infty}}$. 
Table \ref{table:parametersR} summarizes our runs with orbital motion. We ran all the runs of the static case again including orbital motion, and they are designated with the suffix `R'.
Figure \ref{fig:46} shows Run 7R which has $\eta=0.005$ and includes orbital motion.
In intermediate cases (e.g., Run 8R) the cavity of the secondary wind looks different and full of mixed primary and secondary gas instead of mostly separated two winds that mix only close to the conical shell.
Figure \ref{fig:53} shows Run 13R for $\eta=3.125\times10^{-4}$ and orbital motion. This run can be compared with Run 13 in Figure \ref{fig:27} in which the stars are held static.
We can see that the dense line behind the secondary is curved. This has however very little effect on the accretion rate as the orbital velocity is small compared with the wind velocity.
Figure \ref{fig:3panel_mdot_orb} shows the mass accretion rate as a function of time for all the simulation with orbital motion. The results are qualitatively similar to Figure \ref{fig:3panel_mdot}. 
In some cases, especially in some of the higher values of $\eta$ we see that the accretion starts earlier when the orbital motion is added.
For very low $\eta$ the accretion is still continuous though fluctuating more than for the case with static wind.
%
%
%

\section{Discussion}
\label{sec:discussion}

\begin{figure*}
\centering
\includegraphics[trim= 0.0cm 0.0cm 1.0cm 1.5cm,clip=true,width=0.90\textwidth]{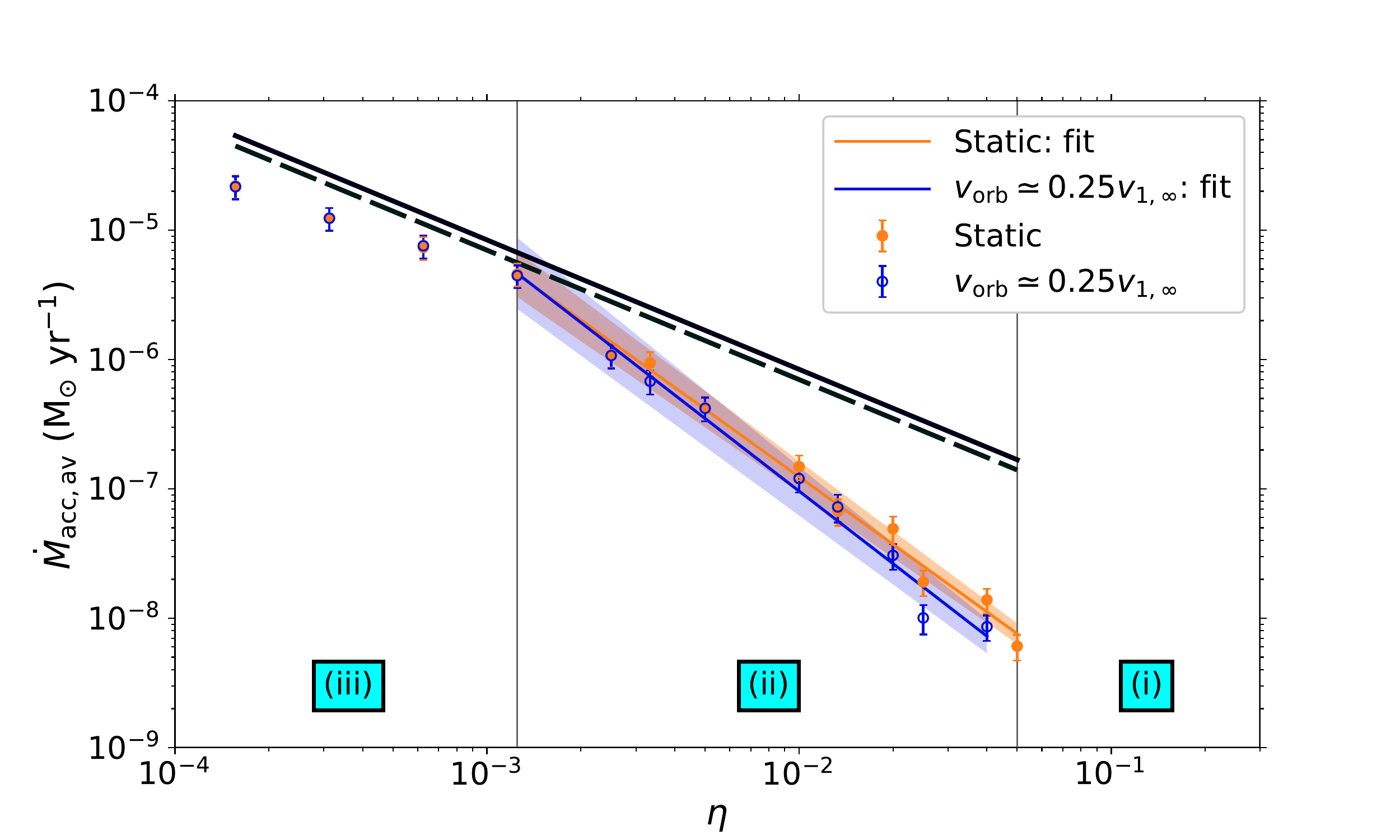} 
\caption{
The $\dot{M}_{\rm acc}$ -- $\eta$ diagram.
The accretion rate obtained from Runs 2--14 and 2R--14R. The error bars indicate 95\% confidence level (and for most points are smaller than the symbols).
Runs 1 and two more runs that we do not label here with $\eta>0.05$ did not yield any accretion.
There are three different regions in the figure (see text). The orange and blue straight line indicated the power law described in equations (\ref{eq:mdotetapowerlaw}) and (\ref{eq:mdotetapowerlaworb}). The shaded lines around the lines indicate the fit $1\sigma$ error.
The solid black line describes the BHL accretion rate as in equation (\ref{eq:MdotBHL}), and the dashed black line shows the BHL when the orbital motion is included.
}
\label{fig:Mdoteta}
\end{figure*}

Figure \ref{fig:Mdoteta} shows the time averaged accretion rate $\dot{M}_{\rm acc, av}$ in our simulations as a function of $\eta$. The average is done from the time accretion starts till the end of the simulation.
Run 1, 1R and two more runs for each case with $\eta>0.05$ that we discussed above and do not label here did not yield any accretion and therefore are not shown in the diagram as they have $\dot{M}_{\rm acc, av}=0$.

For simulations in which the secondary wind is blowing against the primary wind we do not expect the accretion to resemble Bondi-Hoyle-Lyttleton (BHL; \citealt{HoyleLyttleton1939},\citealt{BondiHoyle1944}; see also  a review by \citealt{Edgar2004}) accretion, as the assumptions for the BHL do not hold in the setting of a colliding wind system.
Calibrating for the parameters in our simulations, the BHL accretion rate for static stars is 
\begin{equation}
\begin{split}
\dot{M}_{\rm BHL} &=4.2\times 10^{-8} \left(\frac{M_2}{20 \rmModot}\right)^2 \left(\frac{\dot{M_1}}{3 \times10^{-4} \msyr}\right) \\
&\times  \left(\frac{v_1}{500 \kms}\right)^{-4} \left(\frac{a}{6 \AU}\right)^{-2} \msyr,
\end{split}
\label{eq:MdotBHL}
\end{equation}
and when including orbital motion the rate is a little lower. The rates for both cases are indicated as slant lines in Figure \ref{fig:Mdoteta}, a solid line for the static case and a dashed line for the orbiting case.

We can identify different regions in the $\dot{M}_{\rm acc, av}$--$\eta$ diagram (Figure \ref{fig:Mdoteta}):
\newline
\textbf{(i) No accretion:} For $0.05\lesssim\eta$ the secondary wind pushes away all the primary wind material so there is a well-defined CWS and no accretion ($\dot{M}_{\rm acc, av}=0$).
\newline
\textbf{(ii) Sub-BHL Accretion:}
(iia) For $0.01\lesssim\eta\lesssim0.05$ there is a transition region, in which accretion is very sporadic. Mass can occasionally be accreted but for most of the time the secondary wind and radiation prevent accretion. This is the region where radiative breaking is dominant.
(iib) For $0.001\lesssim\eta\lesssim0.01$ accretion occurs most of the time.
The accretion rate and the accretion duty cycle are larger as $\eta$ decreases. We fit the results of the simulations in region (ii) and find a power-law relation that satisfies
    \begin{equation}
        \dot{M}_{\rm acc, av} \propto \eta^{-1.73\pm0.06} ~(\rm{Static}).
        \label{eq:mdotetapowerlaw}
    \end{equation}
The stated error in the power represents 1$\sigma$ and the line is shown in orange in Figure \ref{fig:Mdoteta}. 
\newline
\textbf{(iii) BHL accretion:} For $\eta\lesssim0.001$ the accretion becomes continuous in time and the accretion rate is BHL. We discuss it further below.

For region (ii) the accretted gas comes mostly from clumps that arrive to the side of the secondary facing the CWS and not from behind the star as for BHL accretion.
The secondary wind interferes with the arrival of gas towards the star and therefore we do not expect it to be described as BHL.
Region (iii) (Runs 11-14) does not describe the colliding wind problem in the sense that there is no situation of two colliding winds with post-shocked gas, and not a CWS as for larger $\eta$.
This is the reason for the change of the trend in the upper panel of Figure \ref{fig:3panel_mdot}.
Our results for Runs 11--14 for the very low values of $\eta$ correspond a strong mass loss rate from the LBV, as occurs during giant eruptions. 

For most of the runs with orbital motion included, the mass accretion rate is slightly smaller than the case without orbital motion, though the change is small enough to fall within the error range.
The three regions described above are that same, and we found similar accretion rates for region (iii), and a slightly lower accretion rate for region (ii).
The power-law fit for region (ii) gives a steeper slope of
    \begin{equation}
        \dot{M}_{\rm acc, av} \propto \eta^{-1.86\pm0.09}  ~(\rm{Orbit}).
        \label{eq:mdotetapowerlaworb}
    \end{equation}
This line is shown in blue in Figure \ref{fig:Mdoteta}.
We note that the 1$\sigma$ errors for the fits overlap.
One small difference we found between the orbital motion and static cases is that for the runs with orbital motion accretion started at a low rate for $\eta=0.04$, slightly higher than for the static case.

The instabilities in the post-shocked secondary wind create small filaments and clumps. Then, gravity of the secondary pulls them together with a push from the primary wind. Against these forces the secondary wind pushes back, and as a result there is an ongoing quasi-stable state of CWS.
Occasionally the filaments or clumps reach the secondary and get accreted, as we quantify in the simulations presented here.
For region (ii) the accretion is intermittent while for region (iii) it is continuous. To quantify this we measure the accretion duty cycle
\begin{equation}
D_{\rm acc}=\frac{t_{\rm acc}}{t_{\rm tot}},
\label{eq:dutycycle}
\end{equation}
where $t_{\rm acc}$ is the duration in which the secondary accrete and $t_{\rm tot}$ is the duration of the run (after accretion starts).

\begin{figure*}
\centering
\includegraphics[trim= 0.0cm 0.0cm 1.0cm 1.5cm,clip=true,width=0.90\textwidth]{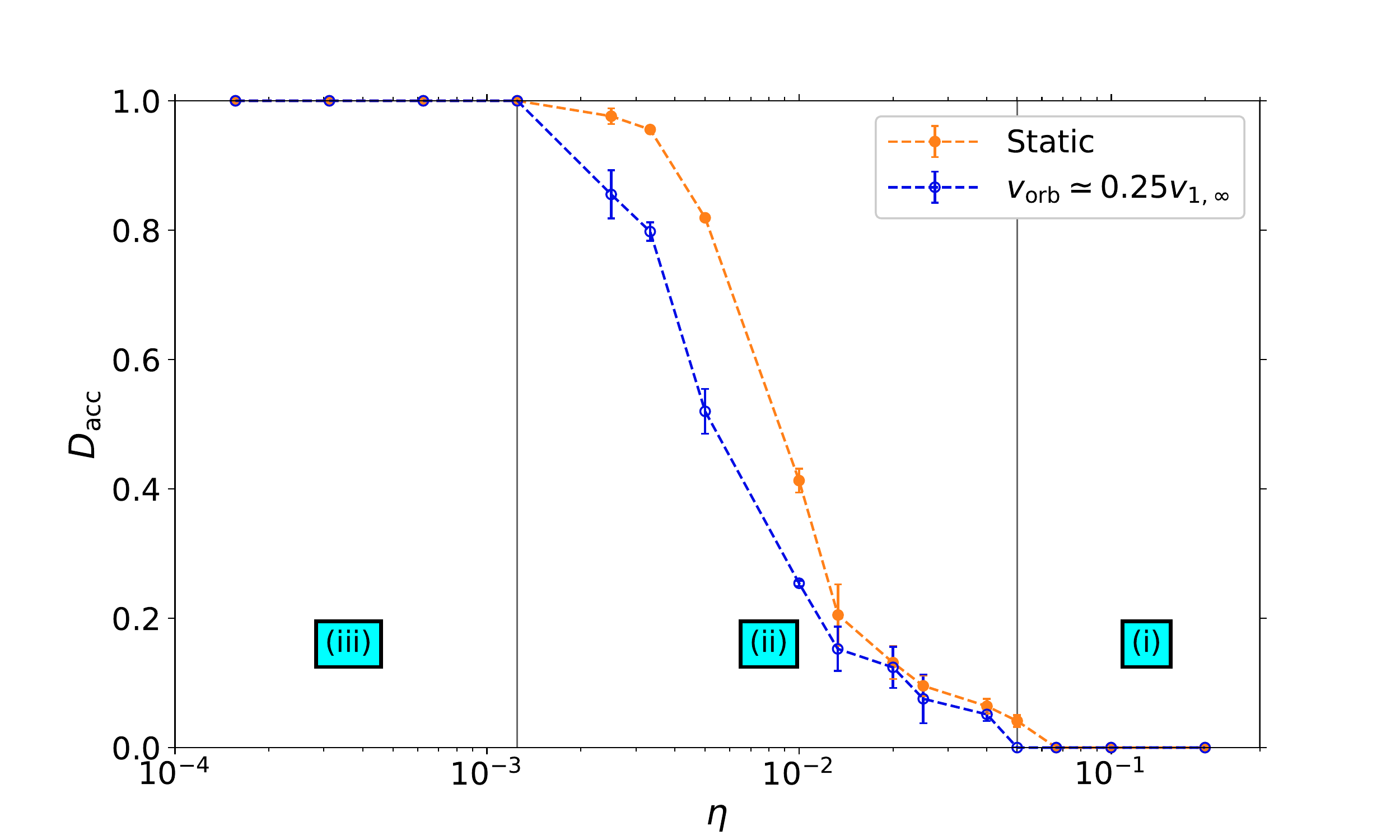} 
\caption{
The accretion duty cycle $D_{\rm acc}$ (Equation \ref{eq:dutycycle}) depending on the momentum ratio $\eta$.
Error bars indicate the difference between calculating the duty cycle for the entire duration of the simulation and half its duration.
The accretion duty cycle is zero in region (i), and then as $\eta$ decreases the curve shows in region (ii) a quite steep transition from 0 to 1 around $\eta\approx0.01$. In region (iii) the accretion happens contiguously. For the case with  orbital motion the duty cycle is lower than for the static case.
The three regions discussed in section \ref{sec:discussion} are marked on the plot.
}
\label{fig:accdutycycleeta}
\end{figure*}

Figure \ref{fig:accdutycycleeta} shows the accretion duty cycle depending on the momentum ratio. The simulations show transition from $D_{\rm acc}=0$ for $\eta>0.05$, to $D_{\rm acc}=1$ for $\eta \lesssim 1.25 \times 10^{-3}$.
The line has a sigmoid shape for both the orbiting and static cases.
The transition is relatively sharp, occurring close to $\eta \approx 0.01$.
When orbital motion is included the duty cycle is lower than for the static case.
%

\begin{figure*}
\centering
\includegraphics[trim= 0.0cm 0.0cm 1.0cm 1.5cm,clip=true,width=0.90\textwidth]{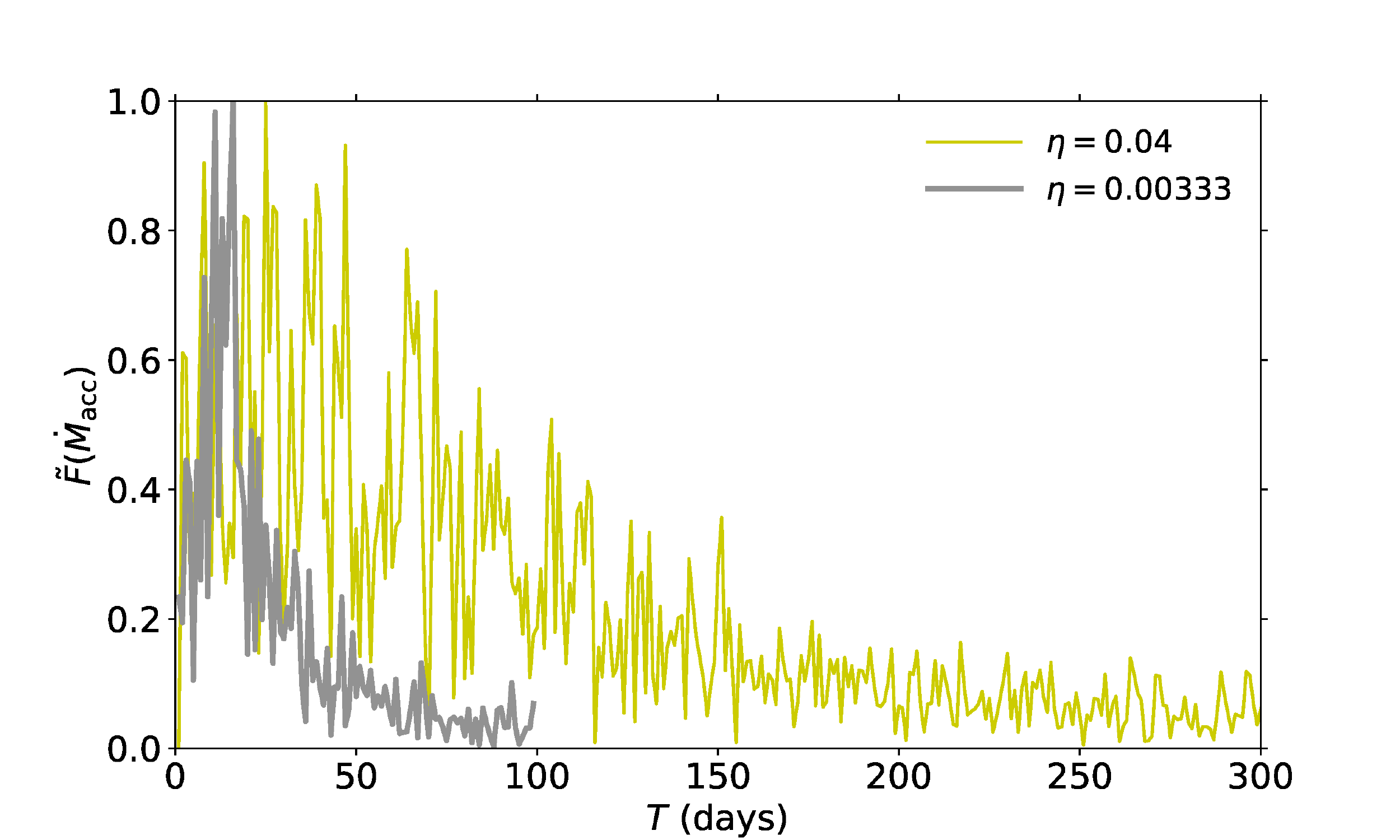} 
\caption{
Fourier transformation for two characteristic accretion-rates in our static simulations. The abscissa shows the accretion period $T$ (the inverse of the accretion frequency), and the ordinate shows the normalized Fourier transform $\tilde{F}(\dot{M}_{\rm acc})$. Run 3 ($\eta=0.04$) and Run 9 ($\eta=3.33\times 10^{-3}$), which are on opposite edges of region (ii).
The time dependent accretion rate for both runs is shown in Figure \ref{fig:3panel_mdot}, though Run 3 was simulated for much longer time than shown.
There are no clear characteristic frequencies, but it is evident that for lower value of $\eta$ accretion occurs much more frequently.
}
\label{fig:Fourier}
\end{figure*}

Running a Fourier analysis for the accretion rate as a function of time we obtain its power spectrum.
The leading frequencies in each run were in the range corresponding to 7--25 days, but by themselves they do not show a clear correlation with $\eta$.
To obtain a characteristic accretion frequency we used a weighted average of the power spectra.
We find that the accretion frequency increases with increasing $\eta$. The average accretion period for $\eta=0.04$ is $\simeq 60 \days$, while for $\eta=3.33\times 10^{-3}$ it is only $\simeq 19 \days$ (Figure \ref{fig:Fourier}). In both cases accretion occurs at lower rates (more frequently). The difficulty in obtaining clear periods is an indication of the accretion process being stochastic, and strongly dependent on random accretion of clumps and the ability of the secondary wind to deflect or destroy them (completely or partially) before being accreted.
%
%

The accretion rate is limited from above by a rate approximately equal to the BHL accretion rate (Equation \ref{eq:MdotBHL}). This is true for both cases, with or without orbital motion.
The BHL accretion radius is
\begin{equation}
R_{\rm acc}=\frac{2GM_2}{v_1^2} \simeq 30.5
\left(\frac{M_2}{20 \rmModot}\right)
\left(\frac{v_1 \vphantom{\rmModot}}{500 \kms}\right)^{-2}
\rmRodot.
\label{eq:R_BHL}
\end{equation}
This value is constant for our simulations as when we change $\eta$ we change only $\dot{M}_1$ but leave $v_1$ the same.
For large enough values of $\eta$, the CWS is outside of $R_{\rm acc}$, as shown in Figure \ref{fig:density_smooth_16}.
When we decrease $\eta$ the CWS, or at least parts of it, can get into $R_{\rm acc}$.

We calculate the value of $\eta$ where the apex of the colliding wind system reaches $R_{\rm acc}$.
If the distance of the apex to the secondary is 
$r_2=a\eta^{1/2}/(1+\eta^{1/2})$
\citep{Usov1992},
then the apex enters the BHL accretion radius when
\begin{equation}
\eta_{\rm BHL}\approx 6 \times 10^{-3}
\left(\frac{M_2}{20 \rmModot}\right)
\left(\frac{v_1 \vphantom{\rmModot}}{500 \kms}\right)^{-2}
\left(\frac{a\vphantom{\rmRodot}}{6 \AU}\right)^{-1}.
\label{eq:eta_BHL}
\end{equation}
Note that this is a crude estimate as (1) only the part close to the apex of the CWS will enter the BHL accretion radius and (2) we have CWS with instabilities rather than a smooth one for which the approximation was applied.
Nevertheless, our estimate for $\eta_{\rm BHL}$ is in the range where the accretion makes the sharp transition towards region (iii), as seen in Figure \ref{fig:accdutycycleeta}.

\begin{figure*}
\centering
\includegraphics[trim= 0.0cm 0.0cm 0.0cm 0.0cm,clip=true,width=0.99\textwidth]{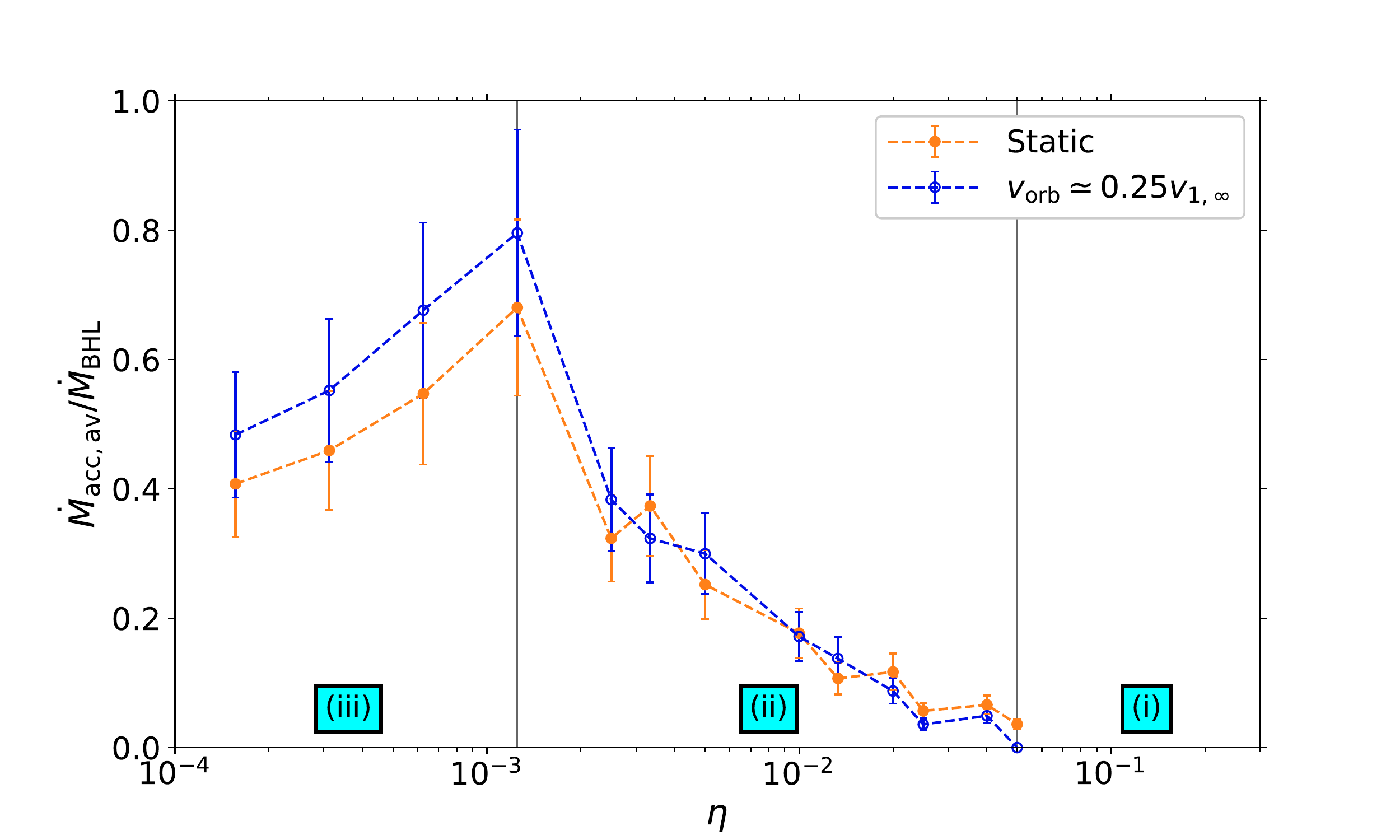} 
\caption{
The mass accretion rate relative to the BHL mass accretion rate. The error bars indicate 95\% confidence level. In region (iii) we expect BHL accretion, but the accretion rate obtained by the simulations is smaller than the classical BHL accretion rate (Equation \ref{eq:MdotBHL}). In region (ii) the accretion rate is sub-BHL. In region (ii) we obtain BHL accretion with a factor of 0.4-0.8, as discussed in section \ref{sec:discussion}.
}
\label{fig:mdot_BHL}
\end{figure*}

The classical BHL rate (Equation \ref{eq:MdotBHL}) is acutely higher than the obtained rate. 
In Figure \ref{fig:mdot_BHL} we plot the fraction of the accretion rate obtained in our simulations over the classical BHL accretion rate. In region (iii) we expect to obtain wind accretion, however the mass accretion rate we obtained is smaller than the BHL accretion rate, in the range
\begin{equation}
\dot{M}_{\rm acc, av} \simeq (0.4-0.8) \dot{M}_{\rm BHL}.
\label{eq:Mdot_BHL_fraction}
\end{equation}
Namely, the accretion in region (iii) is described by BHL accretion but the accretion rate is lower than the classical value.

There are a few sources for the difference. The main difference between our simulation and the classical BHL paradigm is that in our simulations the adiabatic index is different and can be non-constant.
\cite{Livioetal1986} showed that the BHL accretion rate for $\gamma=4/3$ is $0.72\dot{M}_{\rm BHL}$ and for $\gamma=5/3$ is $0.48\dot{M}_{\rm BHL}$.
The value of $\gamma$ in our simulations varies within this range, and indeed we obtained $\dot{M}_{\rm acc}/ \dot{M}_{\rm BHL}$ as expected by the results of \cite{Livioetal1986}. 
In some simulations of accretion during common-envelope evolution even lower accretion rates than \cite{Livioetal1986} were obtained \citep{MacLeodRamirez-Ruiz2015,MacLeodetal2017}, though this was obtained in a geometry and hydrodynamics different than the classical BHL.
Another factor that lowers the accretion rate in our simulations relative to the BHL rate is a difference in the velocity field due to the proximity of the primary. This makes the wind diverge when arriving to the secondary rather than being parallel, resulting in a lower accretion rate.
%

In our simulations we did not modify the stellar atmosphere as a result of accretion. Accretion at a very high rate can lead to a reduction in the effective temperature of the star and in turn weaken its wind, as suggested by \cite{KashiSoker2009} for accretion close to the periastron passage of $\eta$~Carinae. This effect was later obtained in simulations \citep{Kashi2017}. 
What would be the influence of this accretion wind weakening effect on the accretion rate in the present set of simulations?
The primary wind might be able to be accreted more easily as the secondary wind will effectively have lower momentum. The accretion wind weakening effect might therefore shift the results we obtained for the mass accretion rate in the $\dot{M}
_{\rm acc}$ -- $\eta$ diagram (Figure \ref{fig:Mdoteta}) towards a higher value of $\eta$. This requires additional modeling and a detailed investigation of this effect shall be explored in a future paper.

\section{SUMMARY}
\label{sec:summary}

In this paper we describe a numerical experiment of a colliding wind LBV--WR system under different conditions of primary (LBV) mass loss rate, that translates to the momentum ratio $\eta$ (Equation \ref{eq:eta}). We model the accelerating winds, the gravity of the stars, cooling, and radiation transfer.
We follow the simulations to about 200 days (longer for Runs 2--4 in order to get better statistics for the average accretion rate and duty cycle) and measure the accretion onto the secondary star (the WR star).

We plot the average mass accretion rate dependency on the momentum ratio (Figure \ref{fig:Mdoteta}) and identify different regions.
For $\eta>0.05$ there is no accretion (region (i) in Figure \ref{fig:Mdoteta}). For smaller values accretion occurs while the wind of the secondary is still dominant, and we get sub-BHL accretion (region (ii) in Figure \ref{fig:Mdoteta}).
The instabilities in the CWS form clumps that have a velocity component in the direction of the flow (namely, along the sides of the CWS) but also have the acceleration of the secondary. Depending on the parameters, some of the clumps can be pulled towards the secondary and get accreted. 
As long as the momentum ratio is not too small, the wind and radiation of the secondary is able to either  push part of the clumps away, cause braking, or deflect them to prevent them from being accreted. This part becomes smaller as $\eta$ decreases. The result is that
the accretion follows a power law of about $\dot{M}_{\rm acc} \propto \eta^{-1.73}$ for static stars and steeper $\dot{M}_{\rm acc} \propto \eta^{-1.86}$ when orbital motion is included (Equations \ref{eq:mdotetapowerlaw} and \ref{eq:mdotetapowerlaworb}). 

For very low $\eta$ the secondary wind almost does not exist.
Accretion under such conditions (region (iii) in Figure \ref{fig:Mdoteta}) is directly onto the secondary as there is no colliding wind structure, but instead a geometry that resembles BHL accretion, with some differences -- the wind velocity vectors are not parallel and its value is not constant and exposed to the stellar radiation. We obtain an accretion rate of about 0.4--0.8 the BHL accretion rate, in agreement with previous results in the literature \citep{Livioetal1986}.

Though our simulations do not claim to account for all systems and cover the vast parameter space, they give a general quantitative representation of accretion in a massive binary for different values of wind momentum ratio.

The method proposed here -- a systematic exploration of parameter space of colliding wind binaries -- has shown to give quantitative relations between measurable parameters. Expanding this method to the study of other parameters such as distances between the stars, orbital motion, eccentricity, and different wind acceleration schemes and measuring relevant parameters in these simulations will allow us to derive more of such relations.
These relations are going to be helpful when connected to parameters of specific colliding wind binaries. By doing so, we can get better constraints on the stellar wind parameters, and in turn on the massive stars themselves. We hope this paper will open a new window to the study of colliding winds using extensive coverage of parameters in high resolution 3D numerical simulations.

\section*{Acknowledgements}
We thank an anonymous referee for very helpful comments.
We acknowledges support from the R\&D authority in Ariel University.
We acknowledge the Ariel HPC Center at Ariel University for providing computing resources that have contributed to the research results reported within this paper.

\section*{Data Availability}
The data underlying this article will be shared on a reasonable request to the corresponding author.

\label{lastpage}

\begin{thebibliography}{}

\bibitem[Akashi et al.(2013)]{Akashietal2013} Akashi, M.~S., Kashi, A., \& Soker, N.\ 2013, \na, 18, 23.

\bibitem[Bestenlehner et al.(2022)]{Bestenlehneretal2022} Bestenlehner, J.~M., Crowther, P.~A., Broos, P.~S., Pollock, A.~M.~T., \& Townsley, L.~K.\ 2022, \mnras, 510, 6133.


\bibitem[Bondi \& Hoyle(1944)]{BondiHoyle1944} Bondi, H., \& Hoyle, F.\ 1944, \mnras, 104, 273.

\bibitem[Castor, Abbott, \& Klein(1975)]{Castoretal1975} Castor, J.~I., Abbott, D.~C., \& Klein, R.~I.\ 1975, \apj, 195, 157.

\bibitem[Colella \& Woodward(1984)]{ColellaWoodward1984} Colella, P., \& Woodward, P.~R.\ 1984, Journal of Computational Physics, 54, 174 

\bibitem[Crowther(2007)]{Crowther2007} Crowther, P.~A.\ 2007, \araa, 45, 177.

\bibitem[Davidson \& Humphreys(2012)]{DavidsonHumphreys2012} Davidson, K., \& Humphreys, R.~M.\ 2012, Eta Carinae and the Supernova Impostors, 384. 

\bibitem[De Marco \& Izzard(2017)]{DeMarcoIzzard2017} De Marco, O., \& Izzard, R.~G.\ 2017, \pasa, 34, e001 

\bibitem[Eatson et al.(2022)]{Eatsonetal2022} Eatson, J.~W., Pittard, J.~M., \& Van Loo, S.\ 2022, arXiv e-prints, arXiv:2204.12354.

\bibitem[Edgar(2004)]{Edgar2004} Edgar, R.\ 2004, \nar, 48, 843.

\bibitem[Eichler \& Usov(1993)]{EichlerUsov1993} Eichler, D., \& Usov, V.\ 1993, \apj, 402, 271.

\bibitem[Eldridge(2017)]{Eldridge2017} Eldridge, J.~J.\ 2017, Handbook of Supernovae, ISBN 978-3-319-21845-8.~Springer International Publishing AG, 2017, 671

\bibitem[Eldridge et al.(2008)]{Eldridgeetal2008} Eldridge, J.~J., Izzard, R.~G., \& Tout, C.~A.\ 2008, \mnras, 384, 1109 

\bibitem[Eldridge et al.(2015)]{Eldridgeetal2015} Eldridge, J.~J., McClelland, L.~A.~S., Xiao, L., Stanway, E.~R., \& Bray, J.\ 2015, Wolf-Rayet Stars, 177.

\bibitem[Eldridge \& Stanway(2022)]{EldridgeStanway2022} Eldridge, J.~J., \& Stanway, E.~R.\ 2022, arXiv e-prints, arXiv:2202.01413.

\bibitem[Farrell, Groh, Meynet, \& Eldridge(2022)]{Farrelletal2022} Farrell, E., Groh, J.~H., Meynet, G., \& Eldridge, J.~J.\ 2022, \mnras, 512, 4116.

\bibitem[Fryxell et al.(2000)]{Fryxell2000} Fryxell, B., Olson, K., Ricker, P., et al.\ 2000, \apjs, 131, 273

\bibitem[Folini \& Walder(2000)]{FoliniWalder2000} Folini, D., \& Walder, R.\ 2000, \apss, 274, 189 

\bibitem[Folini \& Walder(2002)]{FoliniWalder2002} Folini, D., \& Walder, R.\ 2002, Interacting Winds from Massive Stars, 260, 605 

\bibitem[Garofali et al.(2019)]{Garofalietal2019} Garofali, K., Levesque, E.~M., Massey, P., \& Williams, B.~F.\ 2019, \apj, 880, 8.

\bibitem[Georgy et al.(2017)]{Georgyetal2017} Georgy, C., Hirschi, R., \& Ekstr{\"o}m, S.\ 2017, Second BRITE-Constellation Science Conference: Small Satellites - Big Science, 5, 37.

\bibitem[Hainich et al.(2014)]{Hainichetal2014} Hainich, R., R{\"u}hling, U., Todt, H., et al.\ 2014, \aap, 565, A27.


\bibitem[Heger et al.(2000)]{Hegeretal2000} Heger, A., Langer, N., \& Woosley, S.~E.\ 2000, \apj, 528, 368 

\bibitem[Hoyle \& Lyttleton(1939)]{HoyleLyttleton1939} Hoyle, F., \& Lyttleton, R.~A.\ 1939, Proceedings of the Cambridge Philosophical Society, 35, 405.


\bibitem[Hillier et al.(2019)]{Hillieretal2019} Hillier, D.~J., Koenigsberger, G., Naz{\'e}, Y., et al.\ 2019, \mnras, 486, 725.

\bibitem[Iglesias \& Rogers(1996)]{IglesiasRogers1996} Iglesias, C.~A., \& Rogers, F.~J.\ 1996, \apj, 464, 943.


\bibitem[Ishii et al.(1993)]{Ishiietal1993} Ishii, T., Matsuda, T., Shima, E., et al.\ 1993, \apj, 404, 706.

\bibitem[Kashi(2017)]{Kashi2017} Kashi, A.\ 2017, \mnras, 464, 775

\bibitem[Kashi(2019)]{Kashi2019} Kashi, A.\ 2019, \mnras, 486, 926.

\bibitem[Kashi(2020)]{Kashi2020} Kashi, A.\ 2020, \mnras, 492, 5261.

\bibitem[Kashi et al.(2016)]{Kashietal2016} Kashi, A., Davidson, K., \& Humphreys, R.~M.\ 2016, \apj, 817, 66 

\bibitem[Kashi \& Michaelis(2021)]{KashiMichaelis2021} Kashi, A., \& Michaelis, A.\ 2021, Galaxies, 10, 4.

\bibitem[Kashi \& Soker(2009)]{KashiSoker2009} Kashi, A., \& Soker, N.\ 2009, \na, 14, 11.

\bibitem[Kobulnicky, Chick, \& Povich(2019)]{Kobulnickyetal2019} Kobulnicky, et al.\ 2019, \aj, 158, 73.

\bibitem[L{\'e}pine \& Moffat(1999)]{LepineMoffat1999} L{\'e}pine, S., \& Moffat, A.~F.~J.\ 1999, \apj, 514, 909.

\bibitem[Marcolino et al.(2009)]{Marcolinoetal2009} Marcolino, W.~L.~F., Bouret, J.-C., Martins, F., et al.\ 2009, \aap, 498, 837.

\bibitem[Koenigsberger et al.(2014)]{Koenigsbergeretal2014}
Koenigsberger, G., Morrell, N., Hillier, D.~J., et al.\ 2014, \aj, 148, 62.

\bibitem[Kudritzki \& Puls(2000)]{KudritzkiPuls2000} Kudritzki, R.-P., \& Puls, J.\ 2000, \araa, 38, 613 

\bibitem[Langer(2012)]{Langer2012} Langer, N.\ 2012, \araa, 50, 107 

\bibitem[Leitherer(1997)]{Leithereretal1997} Leitherer, C.\ 1997, Luminous Blue Variables: Massive Stars in Transition, 120, 58.

\bibitem[Livio et al.(1986)]{Livioetal1986} Livio, M., Soker, N., de Kool, M., \& Savonije, G.~J.\ 1986, \mnras, 222, 235.

\bibitem[MacLeod et al.(2017)]{MacLeodetal2017} MacLeod, M., Antoni, A., Murguia-Berthier, A., Macias, P., \& Ramirez-Ruiz, E.\ 2017, \apj, 838, 56.

\bibitem[MacLeod \& Ramirez-Ruiz(2015)]{MacLeodRamirez-Ruiz2015} MacLeod, M., \& Ramirez-Ruiz, E.\ 2015, \apjl, 798, L19.

\bibitem[Maeder(2009)]{Maeder2009} Maeder, A.\ 2009, Astronomy and Astrophysics Library.~ISBN 978-3-540-76948-4.~Springer Berlin Heidelberg, 2009  

\bibitem[Mahy et al.(2022)]{Mahyetal2022} Mahy, L., Lanthermann, C., Hutsem{\'e}kers, D., et al.\ 2022, \aap, 657, A4.

\bibitem[Mason et al.(2009)]{Masonetal2009} Mason, B.~D., Hartkopf, W.~I., Gies, D.~R., Henry, T.~J., \& Helsel, J.~W.\ 2009, \aj, 137, 3358 

\bibitem[McLeod \& Whitworth(2013)]{McLeodWhitworth2013} McLeod, A.~D., \& Whitworth, A.~P.\ 2013, \mnras, 431, 710.

\bibitem[M{\"u}ller \& Vink(2008)]{MullerVink2008} M{\"u}ller, P.~E., \& Vink, J.~S.\ 2008, \aap, 492, 493

\bibitem[Nagae et al.(2004)]{Nagaeetal2004} Nagae, T., Oka, K., Matsuda, T., et al.\ 2004, \aap, 419, 335.

\bibitem[Naz{\'e} et al.(2018)]{Nazeetal2018} Naz{\'e}, Y., Koenigsberger, G., Pittard, J.~M., et al.\ 2018, \apj, 853, 164.



\bibitem[Nugis \& Lamers(2000)]{NugisLamers2000} Nugis, T., \& Lamers, H.~J.~G.~L.~M.\ 2000, \aap, 360, 227.

\bibitem[Oskinova et al.(2007)]{Oskinovaetal2007} Oskinova, L.~M., Hamann, W.-R., \& Feldmeier, A.\ 2007, \aap, 476, 1331.

\bibitem[Owocki(2010)]{Owocki2010} Owocki, S.\ 2010, Hot and Cool: Bridging Gaps in Massive Star Evolution, 425, 199.

\bibitem[Owocki(2015)]{Owockietal2015} Owocki, S.~P.\ 2015, Very Massive Stars in the Local Universe, 412, 113.

\bibitem[Owocki \& Rybicki(1984)]{OwockiRybicki1984} Owocki, S.~P., \& Rybicki, G.~B.\ 1984, \apj, 284, 337.

\bibitem[Pauldrach et al.(1986)]{Pauldrachetal1986} Pauldrach, A., Puls, J., \& Kudritzki, R.~P.\ 1986, \aap, 164, 86.

\bibitem[Pejcha et al.(2022)]{Pejchaetal2022} Pejcha, O., Calder{\'o}n, D., \& Kurf{\"u}rst, P.\ 2022, \mnras, 510, 3276.

\bibitem[Pollock et al.(2018)]{Pollocketal2018} Pollock, A.~M.~T., Crowther, P.~A., Tehrani, K., Broos, P.~S., \& Townsley, L.~K.\ 2018, \mnras, 474, 3228.


\bibitem[Puls et al.(2006)]{Pulsetal2006} Puls, J., Markova, N., Scuderi, S., et al.\ 2006, \aap, 454, 625.


\bibitem[Puls et al.(2008)]{Pulsetal2008} Puls, J., Vink, J.~S., \& Najarro, F.\ 2008, \aapr, 16, 209 

\bibitem[Quataert et al.(2016)]{Quataertetal2016} Quataert, E., Fern{\'a}ndez, R., Kasen, D., Klion, H., \& Paxton, B.\ 2016, \mnras, 458, 1214.

\bibitem[Ruffert(1994)]{Ruffertetal1994} Ruffert, M.\ 1994, \apj, 427, 342.

\bibitem[Sana et al.(2012)]{Sanaetal2012} Sana, H., de Mink, S.~E., de Koter, A., et al.\ 2012, Science, 337, 444 

\bibitem[Schr{\o}der et al.(2021)]{Schroderetal2021} Schr{\o}der, S.~L., MacLeod, M., Ramirez-Ruiz, E., et al.\ 2021, arXiv e-prints, arXiv:2107.09675.

\bibitem[Shenar et al.(2021)]{Shenaretal2021} Shenar, T., Sana, H., Marchant, P., et al.\ 2021, \aap, 650, A147.

\bibitem[Smartt(2009)]{Smartt2009} Smartt, S.~J.\ 2009, \araa, 47, 63.

\bibitem[Smith(2014)]{Smith2014} Smith, N.\ 2014, \araa, 52, 487 

\bibitem[Smith \& Tombleson(2015)]{SmithTombleson2015} Smith, N., \& Tombleson, R.\ 2015, \mnras, 447, 598.

\bibitem[Soker et al.(1986)]{Sokeretal1986} Soker, N., Livio, M., de Kool, M., \& Savonije, G.~J.\ 1986, \mnras, 221, 445.

\bibitem[Stevens et al.(1992)]{Stevensetal1992} Stevens, I.~R., Blondin, J.~M., \& Pollock, A.~M.~T.\ 1992, \apj, 386, 265.

\bibitem[Sundqvist et al.(2010)]{Sundqvistetal2010} Sundqvist, J.~O., Puls, J., \& Feldmeier, A.\ 2010, \aap, 510, A11.

\bibitem[Sutherland \& Dopita(1993)]{SutherlandDopita1993} Sutherland, R.~S., \& Dopita, M.~A.\ 1993, \apjs, 88, 253 

\bibitem[Tehrani et al.(2019)]{Tehranietal2019} Tehrani, K.~A., Crowther, P.~A., Bestenlehner, J.~M., et al.\ 2019, \mnras, 484, 2692.

\bibitem[Usov(1992)]{Usov1992} Usov, V.~V.\ 1992, \apj, 389, 635 

\bibitem[van Marle et al.(2008)]{vanMarleetal2008} van Marle, A.~J., Owocki, S.~P., \& Shaviv, N.~J.\ 2008, First Stars III, 990, 250.

\bibitem[van Marle et al.(2009)]{vanMarleetal2009} van Marle, A.~J., Owocki, S.~P., \& Shaviv, N.~J.\ 2009, \mnras, 394, 595.

\bibitem[Vink(2015)]{Vink2015} Vink, J.~S.\ 2015, Very Massive Stars in the Local Universe, 412, 77 

\bibitem[Vishniac(1994)]{Vishniac1994} Vishniac, E.~T.\ 1994, \apj, 428, 186.

\bibitem[Walder \& Folini(2000)]{WalderFolini2000} Walder, R., \& Folini, D.\ 2000, \apss, 274, 343 

\bibitem[Walder \& Folini(2002)]{WalderFolini2002} Walder, R., \& Folini, D.\ 2002, Interacting Winds from Massive Stars, 260, 595 

\bibitem[Walder \& Folini(2003)]{WalderFolini2003} Walder, R., \& Folini, D.\ 2003, A Massive Star Odyssey: From Main Sequence to Supernova, 212, 139 

\bibitem[Weis \& Bomans(2020)]{WeiseBomans2020} Weis, K., \& Bomans, D.~J.\ 2020, Galaxies, 8, 20.

\bibitem[Williams et al.(2021)]{Williamsetal2021} Williams, P.~M., Morrell, N.~I., Boutsia, K., \& Massey, P.\ 2021, \mnras, 505, 5029.

\bibitem[Zapartas et al.(2021)]{Zapartasetal2021} Zapartas, E., de Mink, S.~E., Justham, S., et al.\ 2021, \aap, 645, A6.

\bibitem[Zhekov(2021)]{Zhekov2021} Zhekov, S.~A.\ 2021, \mnras, 500, 4837.









\end{thebibliography}
\end{document}